# Evaluating satellite and reanalysis rainfall estimates for climate services in agriculture: a comprehensive methodology

Danny Parsons* [1, 2, 3], David Stern [3], Denis Ndanguza [1], Mouhamadou Bamba Sylla [2], James Musyoka [3], John Bagiliko [1, 2], Graham Clarkson [4] and Peter Dorward [4]

[1] Department of Mathematics, School of Science, College of Science and Technology, University of Rwanda, Kigali P.O. Box 3900, Rwanda

[2] African Institute for Mathematical Sciences (AIMS), AIMS Research and Innovation Center, Rue KG590 ST, Kigali, Rwanda

[3] IDEMS International, Reading RG2 7AX, UK

[4] School of Agriculture, Policy and Development, University of Reading, Reading RG6 6BZ , UK

* Corresponding author, danny@aims.ac.za

**Abstract**

*High-resolution rainfall estimates from satellite and reanalysis data sources (SRE) could play a major role in improving climate services for agriculture. This is particularly relevant in regions that predominantly rely on rain-fed farming but lack a dense network of ground-based measurements to provide localised historical climate information, as in most of the Global South. However, there is a need for a framework which practitioners can use to determine the suitability of these estimated data for specific agricultural applications. This paper presents a comprehensive methodology for evaluating the ability of SRE to provide historical rainfall information for agricultural applications, primarily through comparison with ground-based measurements. The methodology comprises five main steps: data selection and pre-processing, spatial and temporal consistency checks, quantitative SRE-gauge comparisons, bias correction, and application specific summaries. The methodology makes use of graphical summaries, standard comparison metrics, and Markov chain models. We describe how users can apply this methodology to evaluate rainfall estimates for specific applications, complementing existing validation studies. Evaluation cases are presented to demonstrate the methodology using five widely used satellite and reanalysis rainfall products and ground-based measurements from 12 stations in Africa and the Caribbean. The case studies demonstrate how the methodology can be applied to examine multiple aspects of the rainfall estimates. While previous validation studies ask "Does the SRE estimate the true rainfall well?", this methodology provides means of establishing "To what extent can an SRE be used for this specific purpose?" and a comprehensive framework for achieving this. This meets a major need for location specific rainfall information to improve climate information services for millions of small-holder farming households.*

# 1 Introduction

Detailed information about historical rainfall is a critical component of climate services for agriculture. This information is particularly important in areas of the world where most of the population depend on rain-fed farming, such as sub-Saharan Africa (OECD and Food Agriculture Organization of the United Nations, 2016). In these areas, rainfall patterns and variability greatly affect agricultural outcomes, such as crop yields (Kashaigili, et al., 2014; Kyei-Mensah, et al., 2019; Declan, et al., 2005). Long-term records of historical rainfall data help in understanding the climatology of the region and in identifying the suitability of different crops, varieties and other farming strategies. Long term records also enable understanding of variability, which is needed to accurately estimate the risks associated with agricultural decision making.

Climate information needs to be localised to best support decision making. The Food and Agriculture Organization of the United Nations (FAO) promotes the concept of "Localized climate services" and recognises that "Climate impacts are location specific" (Food and Agriculture Organization of the United Nations, 2012). This concept presents a challenge in many parts of the world, particularly in sub-Saharan Africa, where the network of ground stations is sparse (World Meteorological Organization, 2020), and largely clustered around areas of dense population (Dinku, 2019). Hence, the spatial coverage is often inadequate which is challenging for providing accurate, local climate information.

Gridded satellite and reanalysis rainfall estimates (SRE) offer a possible solution to this challenge. These estimates of daily rainfall values are based on satellite observations or reanalysis, with high spatial resolution and wide coverage. Several SRE have provided historical rainfall estimates for more than 30 years up to the present (Maidment, et al., 2017; Funk, et al., 2015; Hersbach, et al., 2020; Boogaard, et al., 2020).

Many studies have validated SRE across geographical regions and time periods (see Background). Studies have shown good results on general rainfall properties, such as monthly and seasonal totals. However, relevant rainfall information for agriculture is often based on complex summaries. Assessing the appropriateness of SRE for such summaries requires investigating features of the data, including the occurrence and amounts of daily rainfall, seasonal patterns and the distribution and occurrence of dry spells. The start of the rainy season is one example, where its definition can include a condition on rainfall amount over several days and a check on dry spells. We therefore suggest tailored and specific evaluation methods that encompass these complex components in addition to general comparison studies. These methods can provide further insight into the appropriateness of SRE for climate services in agriculture.

This paper proposes a comprehensive methodology for assessing the potential of SRE to provide historical rainfall information for climate services for agriculture. We focus on applications that address questions often posed by farmers, such as: "When can I plant?", "When does the rainy season start?", "How long is the rainy season?" and "How much rain is there in the rainy season?" (The Food and Agriculture Organization of the United Nations, 2019; Coulibaly, et al., 2015; Clarkson, et al., 2019). The proposed methodology uses comparison techniques often found in validation studies, and new methods specifically tailored for agricultural applications. The methodology should serve as a tool for potential users of SRE to carry out their own evaluations, as well as providing a template for tailoring evaluation for other uses.

Section 2 gives the background to SRE, reviews existing validation studies and describes relevant applications in climate services for agriculture. The proposed methodology is in Section 3. Case studies with example data are in Section 4. Section 5 discusses the value of the proposed methodology and identifies further work.

# 2 Background

## 2.1 Satellite and reanalysis rainfall estimates

The availability of long-term data from meteorological satellites and reanalysis techniques has led to the development of freely available historical estimates of rainfall with high spatial and temporal resolution. The data are gridded and cover large regions of the Earth (Maidment, et al., 2017; Funk, et al., 2015; Hersbach, et al., 2020). Data are available on a daily (Maidment, et al., 2017; Funk, et al., 2015) or hourly basis (Hersbach, et al., 2020), with few or no missing values and for more than 30 years. Hence, these data complement existing rain gauge networks and provide information in locations not served by existing networks.

However, differences between SRE and rain gauge data should be carefully considered and understood before using one in place of the other for specific applications (Tozer, et al., 2012; Ageet, et al., 2022). Producing SRE often involves data smoothing (Tozer, et al., 2012) which generally reduces variability and leads to differences between average rainfall over the pixel area and a point-based measurement from a rain gauge. Reduced variability in rainfall data could lead to underestimation of extreme events, such as heavy rainfall and drought, and therefore also cause an underestimation of risks. Features such as rain day occurrence and extreme rainfall events can have different properties (King, et al., 2013; Tozer, et al., 2012). However, smoothing data spatially and/or temporally will improve the skill of SRE, and hence the choice of temporal and spatial scale to use SRE data at is also a compromise between adequate representation and required skill. Differences might also be due to the algorithm used in constructing rainfall estimates; for example, satellite based estimates are often based on the relationship between cloud-top temperature and rainfall, which is known to have limitations in detecting rainfall from warm clouds (Young, et al., 2014). Hence, there is a need to evaluate SRE to understand their strengths, weaknesses, and differences between rain gauge data to assess their appropriateness before use in a specific application that has previously relied on rain gauge data. Performance can also vary spatially, particularly in complex terrains (Dinku, et al., 2010) and therefore it is important to understand performance at the intended location of use.

## 2.2 Validating SRE

A large body of studies exists validating SRE across geographical regions, typically comparing SRE with gauge-based observations. Gauge data are often “gridded” using spatial interpolation e.g. kriging (Maidment, et al., 2017; Dinku, et al., 2010; Dinku, et al., 2018) for a more like-for-like comparison (i.e. pixel-to-pixel comparisons). For example, Dinku (2010) evaluated seven SRE over Colombia in comparison with a network of 600 rain gauges. The SRE were reasonably good given the complex terrain of the region, although with substantial differences in some areas. Dinku (2018) evaluated the Climate Hazards Center InfraRed Precipitation (CHIRP) and Climate Hazards Center Infrared Precipitation with Station data (CHIRPS) datasets over East Africa in comparison with a merged product that included data from 1200 rain gauges. CHIRP and CHIRPS performed better than the African Rainfall Climatology (ARC2) and particularly well over decadal and monthly time scales. Performance was worse over mountainous and coastal regions. Funk (2015) evaluated the CHIRP and CHIRPS datasets against the Global Precipitation Climatology Centre (GPCC) gridded data product for the wettest three months of the year to assess its usefulness for agricultural drought monitoring. CHIRPS, and to a lesser extent, CHIRP, showed high correlation and low mean absolute error in many parts of the world.

Validation studies have also compared SRE directly with gauge data (i.e. pixel-to-point comparisons) (Ageet, et al., 2022; Katsekpor, et al., 2024). Funk (2015) also validated the CHIRPS data using pixel-to-point comparisons for seasonal rainfall totals across several countries and showed high correlation with rain gauge measurements. Owusu et al. (2019) showed only moderately positive correlations and large biases on a daily time step over Ghana for three SRE, and stronger correlations on monthly, seasonal, and yearly time steps, although still with some biases. Diem et al. (2014) showed that, in pixel-to-point comparisons of three SRE over Western Uganda, some SRE performed well at estimating seasonal totals, but all overestimated rainfall days and were only "borderline satisfactory" at the best locations.

Pixel-to-pixel comparisons provide an assessment of average climate conditions over the pixel area, whereas pixel-to-point comparisons assess SRE's ability to represent the conditions and variability of a point-based measurement. Pixel-to-point comparisons often have poorer results, as described above, but are important to carry out for applications where SRE will be used to represent the climate at a point location, particularly to understand differences in variability. Both types of comparisons are useful and the choice of comparison approach should depend on the intended application.

This growing evidence base validating various SRE is important in showing how general properties of rainfall are well captured by SRE, such as monthly or seasonal rainfall totals. These studies are also useful in indicating areas where algorithms can be improved, such as over certain geographical features, or on daily time scales. However, the requirements of SRE vary depending on how the data will be used and the application it will be used for. General validation studies cannot evaluate the appropriateness of SRE for all types of applications, but they can enable further studies to build on them and include more comprehensive and specific validation. Hence specific validation is also required, which builds on the existing validation evidence base, but evaluates components systematically and allows for focusing on the requirements of a particular application. This methodology follows the suggestion of Tozer et al. (2012) to quantify differences between SRE and understand the implications for a particular use. A fundamental difference between the general evaluation methods and the methodology proposed in this paper is the main question that they try to address. General evaluation methods ask: "How good are the SRE at estimating rainfall?" whereas our proposed methodology tackles the question: "Can SRE be used for this specific purpose, and if so, how, and what needs to be considered?".

### 2.3 Relevant applications

Many applications currently use information from historical rain gauge data. Applications include climate risk management and climate change adaptation at local and community levels where local variability and trends need to be understood (Hellmuth, et al., 2007; Brönnimann, et al., 2019; McGregor, 2015), hydrological applications such as water supply management, and irrigation scheduling (Amarasinghe, et al., 2016; Chen, et al., 2020), which can also be influenced by when the rainy season starts and ends and agricultural applications such as crop simulation modelling (Holzworth, et al., 2014) and weather index insurance (Shirsath, et al., 2019) which require accurate information on extreme rainfall events and droughts.

One application motivating the development of this methodology is the Participatory Integrated Climate Services for Agriculture (PICSA) approach (Clarkson, et al., 2019) which has been used in over 30 countries to date. In the PICSA process, farmers interpret graphs showing events derived from the historical daily rainfall data from a station close to their location. These events include the rainy season start and end dates, season length, total rainfall within the season, rain days per season, and the length of long dry spells. These graphs enable farmers to discuss climate variability, trends and climate change and the risks related to different agricultural options. The overall aims are "to help facilitate locally-specific evidence-based decision-making by farmers" (Clarkson, et al., 2019). Consequently, the climate information presented can significantly affect the livelihoods of the farmers involved and their communities.

The methodology proposed in this paper is designed to be applicable generally to agricultural applications that rely on local climate information from historical rainfall data.

# 3 Proposed methodology

## 3.1 Overview

The proposed methodology provides a framework for assessing the potential of SRE in providing climate services for agriculture. It draws on work from many previous studies, using widely used analyses and metrics. It also introducing new methods that can offer valuable insights into characteristics that are often overlooked in validation studies.

The methodology follows the principle of purpose driven evaluation – recognising that different applications have different criteria that define good performance and therefore evaluation of SRE should be tailored to the specific application. Application specific evaluation of SRE has been employed in hydrological modelling (Gebrechorkos, et al., 2024; Jiang, 2024), drought monitoring and long term rainfall characteristics (Wei, et al., 2021; Bayissa, et al., 2017; Houngnibo, et al., 2023), as well as for specific agricultural applications such as crop growth simulation (Omondi, et al., 2021) and crop modelling (Ovando, et al., 2018; Ramarohetra, et al., 2013). Common to these approaches is the recognition of the importance of understanding how errors in SRE propagate through to the specific uses of SRE. Application specific uses of SRE may include tailored agricultural summaries and indices or outputs from models.

This methodology builds upon the concepts of general validation combined with application specific evaluation and formalises them into a comprehensive five-step methodology. Its design is such that it begins with broad, generalised analyses and becomes increasingly application specific as users progress through the steps. This design means that earlier steps consist of standard, well documented analyses with clear interpretation guidelines while the later steps require greater application knowledge and technical expertise for decision making.

The steps are designed to be implemented sequentially with each step being informed by the results of the previous steps, allowing for progressive evaluation of the SRE. Below is a brief description of each step.

Step 0 (Data selection and pre-processing) is a pre-analysis preparation step. It involves identifying suitable station data and SRE products, determining the temporal scale suitable for the intended application, and determining the spatial scale for comparison with gauge data (pixel-to-pixel or pixel-to-point). These preparatory decisions ensure that subsequent evaluation steps are based on appropriate, high-quality inputs and are properly aligned with the agricultural applications being assessed.

Step 1 (Spatial and temporal consistency of SRE) is the first analysis step of the methodology. It involves analysis of the SRE products separately before comparison with gauge data. Long term averages of annual rainfall totals, number of rain days, and mean rainfall per rainy day are mapped to assess spatial patterns and identify any spatial inconsistencies. Temporal behaviour is evaluated through time series plots of annual or seasonal totals from individual SRE pixels and statistical tests to identify any abrupt shifts or artificial trends that may be present.

Step 2 (Quantitative SRE-gauge comparisons) involves analysis of SRE products in estimating general rainfall characteristics important for agricultural decision making through comparison with gauge data. These characteristics include annual and monthly summaries of rainfall, seasonality, spell persistence, rainfall amount categories and extreme events. The performance across these general characteristics provide an initial understanding of the suitability of SRE for agricultural applications.

Step 3 (Bias correction of SRE) involves considering whether it is necessary for bias correction to be applied to address systematic biases between SRE and gauge data that have been identified in previous steps. This depends both on the results obtained in the previous step (to identify any biases) and the specific application (whether identified biases will impact the end use of the SRE data). This step also involves repeating some comparisons from step 2 to ensure corrections lead to improved alignment with gauge data.

Step 4 (Application specific summaries) involves evaluating the performance of SRE (bias corrected or not) in estimating rainfall summaries directly relevant to the agricultural applications. These include but are not limited to the start and end of rains, and length of the rainy season. Because the definitions of these summaries can vary by location and crop, application knowledge is essential. Application specific outputs from models, such as crop models, may also be included. These checks remain essential even when earlier steps indicate strong quantitative performance. This is an essential step as it provides an opportunity to assess SRE directly on their intended use.

An overview of the five steps of the proposed methodology is provided in Table 1.

Table 1: Overview of the steps and sub-steps of the methodology

| Step | Methods | Tabular/Graphical displays |
|---|---|---|
| ***0. Data selection and pre-processing*** | Gauge data availability assessment and quality control<br>Review and selection of SRE products, their documentation and relevant literature.<br>Determination of temporal scales for evaluation<br>Selection of comparison approach. | Data inventory plots and quality control visualizations.<br>Map of study area showing station locations. |
| ***1. Spatial and temporal consistency of SRE*** | Calculate long term averages of the annual values of the standard summaries per SRE pixel over the location of interest to assess spatial patterns.<br>Apply change-point detection methods to time series of annual values from representative SRE pixels to identify inhomogeneity. | Maps of the long term averages of each summary.<br>Time series plots of annual summaries used to identify inhomogeneity. |
| ***2. Quantitative SRE-gauge comparisons*** | | |
| 2.1 Standard summaries | Calculate annual (and monthly) standard summaries from SRE and gauge data. Calculate chosen metrics between SRE and gauge summaries. | Time series plots of annual/monthly standard summaries for SRE and gauge data together.<br>Tables of metric values. |

| 2.2 Seasonality | Fit zero-order Markov chain model of rainfall occurrence as a proportion for each day of the year for SRE and gauge.<br>Fit multiple models with varying rain day thresholds for SRE. | Line graphs of fitted rainfall occurrence proportions for SRE and gauge data together. |
| --- | --- | --- |
| 2.3 Memory and spell persistence | Fit first-order Markov chain model of rainfall occurrence given previous day state as a proportion for each day of the year for SRE and gauge.<br>Fit multiple models with varying rain day thresholds for SRE. | Line graphs of fitted rainfall occurrence proportions for SRE and gauge data together. |
| 2.4 Rainfall amounts and extremes | Calculate rainfall amounts categories for daily SRE and gauge data and calculate chosen categorical metrics for rain day variable and categories.<br>Calculate extreme wet (e.g. maximum daily rainfall, upper-percentile rainfall, counts of heavy-rainfall days) and dry indicators (e.g., longest dry spell, number of dry spells) usually on an annual basis. | Bar charts/boxplots of metric values for each rain day variable and categories for each location.<br>Time series plots of annual extreme indices. |
| ***3. Bias correction of SRE*** | Consider whether bias correction is needed based on previous results and the specific application requirements.<br>Construct bias corrected daily SRE values using various methods.<br>Recalculate quantitative SRE-gauge comparisons using bias corrected values. | Quantitative SRE-gauge comparisons graphics with bias corrected values. |
| ***4. Application specific summaries*** | Calculate annual values of specific summaries from SRE and gauge data.<br>Calculate continuous and categorical metrics, as appropriate, between SRE and gauge summaries. | Time series plots of annual specific summaries for SRE and gauge data together. |

## 3.2 Performance metrics for evaluating SRE

A wide range of metrics have been used in studies to evaluate rainfall estimates. Each metric answers a different question about the estimates and highlights different aspects of performance. Knoben et al. (2019) argue for “purpose-dependent evaluation metrics” to be used, based on which aspects of performance are important for the intended application. We follow this principle with our methodology and, therefore, the choice of metrics should depend on the aspects of performance relevant to the application.

The choice of metrics also depends on the measures being evaluated and their temporal scale. For example, if working at the daily scale, wet and dry day detection through rainfall occurrence is likely to be important. Rainfall occurrence is a binary variable, hence requires metrics suitable for categorical variables, whereas applications that work on rainfall totals only at a coarser resolution (e.g. pentadal, decadal, monthly) may not require evaluation of daily rainfall occurrence, and would instead focus on metrics for continuous variables.

In the subsections that follow, we outline an overview of commonly used metrics for evaluating (i) continuous rainfall variables and (ii) rainfall occurrence and categorical rainfall variables, along with guidance on how to choose and interpret these metrics. These metrics form the basis for evaluating SRE performance in the broader methodological framework. We also give an example of how an application can affect the choice of metrics.

### *3.2.1 Metrics for comparing continuous rainfall variables*

Metrics used for comparing continuous rainfall variables include measures of systematic error, such as mean error (bias), percentage bias, mean absolute error, and (root) mean square error (Dinku, et al., 2007; Toté, et al., 2015); measures of association, such as correlation coefficient (Dinku, et al., 2007; Alijanian, et al., 2017); measures of variability, such as variability and standard deviation ratio (Acharya, et al., 2019); and combined indices, such as the Nash–Sutcliffe efficiency (Diem, et al., 2014; Dembélé & Zwart, 2016) and the (modified) Kling–Gupta efficiency (Acharya, et al., 2019; Beck, et al., 2019; Baez-Villanueva, et al., 2018). Table 2 presents commonly used metrics, appropriate for SRE evaluation, with guidance on what question each metric answers and how to interpret its results.

Table 2: Commonly used metrics for continuous rainfall variables. $s_i$ is the SRE value at time $i$, $o_i$ is the observation gauge data value at time $i$, $N$ is the length of the series and $\sigma_s$ and $\sigma_o$ are the standard deviations of the SRE and gauge data respectively.

| **Metric** | **Formula** | **What it answers** | **Range & Ideal Value** | **Interpretation** |
|---|---|---|---|---|
| Mean error | $\frac{1}{N}\sum_{i=1}^{N} s_i - o_i$ | Does the SRE systematically overestimate or underestimate the observations? | $-\infty$ to $\infty$; ideal = 0 | Positive values indicate systematic overestimation by the SRE, while negative values indicate underestimation. Values close to zero suggest little overall bias. |
| Mean absolute error | $\frac{1}{N}\sum_{i=1}^{N} \lvert s_i - o_i \rvert$ | How large are the SRE's errors on average, regardless of direction? | 0 to $\infty$; ideal = 0 | Smaller values indicate smaller average errors between the SRE and the observations. |

| Root mean square error | $\sqrt{\frac{1}{N}\sum_{i=1}^{N}(s_i - o_i)^2}$ | What is the average magnitude of the SRE's errors? | 0 to ∞; ideal = 0 | Smaller values indicate better agreement. |
|---|---|---|---|---|
| Correlation Coefficient | $\frac{\sum_{i=1}^{N}(s_i - \mu_s)(o_i - \mu_o)}{\sqrt{\sum_{i=1}^{N}(s_i - \mu_s)^2 \sum_{i=1}^{N}(o_i - \mu_o)^2}}$ | How well does the SRE capture the pattern of variability in rainfall? | –1 to 1; ideal = 1 | Higher values mean the SRE closely follows the observed pattern of rainfall over time. Values near zero or negative indicate poor agreement in the timing of rainfall variations |
| Percent Bias | $100 \times \frac{\sum_{i=1}^{N}(s_i - o_i)}{\sum_{i=1}^{N} o_i}$ | What is the overall over or under estimation of the SRE as a percentage? | –∞% to ∞%; ideal = 0% | Positive values mean the SRE tends to overestimate rainfall; negative values mean it tends to underestimate. Values closer to zero indicate better agreement with observed totals |
| Ratio of standard deviations | $\frac{\sigma_s}{\sigma_o}$ | Does the SRE reproduce the variability of the observed data? | 0 to ∞; ideal = 1 | Values >1 mean the SRE is more variable than the observations; values <1 mean it is less variable. Values near 1 indicate that the SRE captures the magnitude of the variability well. |
| Nash–Sutcliffe efficiency | $1 - \frac{\sum_{i=1}^{N}(s_i - o_i)^2}{\sum_{i=1}^{N}(o_i - \sigma_o)}$ | How much better is the SRE than using the observation mean? | –∞ to 1; ideal = 1 | Values close to 1 indicate high predictive skill. Values near zero indicate performance similar to using the observed mean, while negative values |

| | | | | indicate poorer performance than the mean. |
|---|---|---|---|---|
| Modified Kling-Gupta Efficiency (KGE) | $\sqrt{(r-1)^2 + \left(\frac{\sigma_s}{\sigma_o} - 1\right)^2 + \left(\frac{\mu_s}{\mu_o} - 1\right)^2}$ | How well does the SRE simultaneously represent the correlation, bias, and variability of the observed data? | $-\infty$ to 1; ideal = 1 | Values near 1 indicate good overall performance across correlation, bias, and variability. Negative values indicate poor agreement across one or more components |

The metric choices will be different for different applications. For example, for calculating the average risk of crop failure in the PICSA approach, the start and end of the rainy season summaries are used to determine the proportion of years meeting certain crop related criteria. To obtain similar risks from SRE, these summaries need to have a similar mean and variability as those calculated from the gauge data, hence a bias metric (e.g. percentage bias or mean error) and a variability metric (e.g. ratio of standard deviations) would be appropriate for this use case. Correlation is less important in this case, hence the KGE may not be an appropriate metric, because it combines correlation, bias and variability performance into a single metric. However, for identifying individual years of note with farmers, correlation of yearly values is more important than overall mean and variability. Hence the suitability for different applications can be understood by utilising relevant metrics.

### 3.2.2 Metrics for comparing rainfall occurrence and categorical rainfall variables

A number of metrics have been used for evaluating rainfall occurrence of SRE, including probability of detection, frequency bias, false alarm ratio (Toté, et al., 2015; Alijanian, et al., 2017), as well as overall measures, such as accuracy, Heidke skill score and Hanssen and Kuipers discriminant (Sun, et al., 2019; Liu & Key, 2016). Table 3 presents commonly used metrics for rainfall occurrence and categorical rainfall variables, appropriate for SRE evaluation, with guidance on what question each metric answers and how to interpret its results.

These metrics can also be used for multi-category rainfall variables, for example, to evaluate detection of different daily rainfall intensities by categorising into intensity levels (e.g. no rain, light, moderate, heavy, violent rain). This uses a generalised version of the metric formula from 2 to k categories, as shown in Table 3, and the question it answers and interpretation remain similar. Some metrics are calculated per category, as indicated in Table 3, while others are overall measures of accuracy and skill. This also provides another way to examine detection of extreme rainfall events by SRE, via the higher intensity categories, complementing evaluation of extremes using standard indices on an annual or monthly basis.

Table 3: Common metrics for rainfall occurrence and categorical rainfall variables. $N =$ total number of observations. For rainfall occurrence: $H =$ number of hits (SRE = rain & gauge = rain), $M =$ number of misses (SRE = no rain & gauge = rain), $C =$ number of correct negatives (SRE = no rain & gauge = no rain), $F =$ number of false alarms (SRE = rain & gauge = no rain). For rainfall category variables with $k$

categories: $n_{ij}$ = number of cases with SRE category $i$ and observation category $j$, $n_{i.} = \sum_j n_{ij}$, $n_{.j} = \sum_i n_{ij}$.

| **Metric** | **Formula (rainfall occurrence)** | **Formula (rainfall categories)** | **What it answers** | **Range & Ideal value** | **Interpretation** |
|---|---|---|---|---|---|
| Probability of Detection (POD) | $\frac{H}{H+M}$ | $POD_i = \frac{n_{ii}}{n_{.i}}$ per category $i$ | How many of the actual rainfall events were correctly detected? | 0 to 1; Ideal = 1 | Higher values mean the SRE detects more days when rainfall (category) actually occurred. |
| Accuracy | $\frac{H+C}{N}$ | $\frac{\sum_{i=1}^{k} n_{ii}}{N}$ | How often does the SRE correctly classify rainy and non-rainy days? | 0 to 1; Ideal = 1 | Higher values mean the SRE correctly identifies more of both rainy and non-rainy days, or categories of rainfall in general . |
| Frequency Bias (BIAS) | $\frac{H+F}{H+M}$ | $BIAS_i = \frac{\sum_{i=1}^{k} n_{i.}}{N}$ per category $i$ | Does SRE over- or under-predict rainfall events? | 0 to ∞; Ideal = 1 | Values >1 mean the SRE overestimates rainy days/category; values <1 mean it underestimates. |
| False Alarm Ratio (FAR) | $\frac{F}{H+F}$ | $FAR_i = \frac{n_{i.} - n_{ii}}{n_{i.}}$ per category $i$ | Of the predicted rainfall events, how many did *not* occur? | 0 to 1; Ideal = 1 | Higher values mean more predicted rainfall events did not occur. |
| Heidke Skill Score | $\frac{2(H \cdot C - F \cdot M)}{(H+M)(M+C)+(H+F)(F+C)}$ | $\frac{\sum_i n_{ii} - \sum_i \frac{n_{i.}n_{.i}}{N}}{N - \sum_i \frac{n_{i.}n_{.i}}{N}}$ | How well does the product predict the correct rainfall category for each day? | –∞ to 1; Ideal = 1 | Values near 1 mean the SRE correctly assigns days to rainfall categories better than random chance. Values near 0 or negative |

| | | | | | indicate poor skill. |
|---|---|---|---|---|---|
| Hanssen and Kuipers discriminant | $\frac{H}{H+M} - \frac{F}{F+C}$ | $HK_i = \frac{n_{ii}}{n_{\cdot i}} - \frac{n_{i\cdot} - n_{ii}}{N - n_{\cdot i}}$ per category $i$ | How well does the SRE discriminate between the occurrence and non-occurrence of rainfall events, independent of event frequency? | −1 to 1; Ideal value = 1 | Higher values indicate better ability of the SRE to distinguish between rainy and non-rainy days, or rainfall of different categories in general. |

### 3.2.3 Defining Performance

Classifying the performance of SRE as poor, good or excellent by applying universal thresholds on metrics is technically easy to implement but defining meaningful thresholds is extremely challenging in practice. What constitutes good performance depends on the application, as the same metric value may be acceptable in one context but unacceptable in another. However, some efforts have been made to define practical thresholds. Houngnibo, et al. (2023) categorised KGE values > 0.5 as acceptable, > 0.75 as good and > 0.9 as very good or excellent in the context of SRE. These can serve as general benchmarks, but users should be cautious as the suitability of such thresholds depend on the specific measurement, location and application. Universal thresholds are unlikely to be useful in all contexts, although values used in similar contexts can guide these choices.

For application specific summaries, acceptable ranges should be guided primarily by the needs of the application. For example, if the goal is to categorise years into good and bad seasons, users might define an acceptable percentage of misclassified years. For standard summaries, defining application relevant thresholds is harder. This could be achieved by working backwards to link metric values on standard summaries to errors in application specific summaries, although this may be impractical for complex applications.

An alternative to absolute thresholds is to rank SRE. Ranking to define relative performance is widely used in validation studies and avoids the need to define universal thresholds. The simplest ranking approach is to order the SRE from best to worst for each relevant metric and then summarise the rankings, for example by average ranking or by most first place rankings. This can be done within each step or over all metrics. Combined metrics are also commonly used for ranking SRE. For example, Moccia, et al. (2025) applied median KGE value to identify the “best gridded product” at different spatial scales. However these methods give equal weight to all metrics and ignores the magnitude of the difference between them, which can give a distorted picture of relative performance.

More complex ranking systems have been used to address these limitations. For example, Houngnibo, et al. (2023) applied PROMETHEE II, a multi-criteria decision analysis (MCDA) method, to rank SRE to assess suitability for drought monitoring, basin water resources assessments, and long term rainfall characteristics, using different configurations for each application. This method allows assigning different weights to metrics, defining a preference and indifference threshold and a preference function, which together describe how differences in metric values should translate into preference.

See (Houngnibo, et al., 2023) for an example of these choices for common metrics and how different criteria are used for different applications. These configurations enable highly customised rankings tailored to specific applications. However, they requires technical expertise to define the configuration and implement them, and results may be sensitive to these choices. Simpler variations of this method, such as normalising metrics to a 0-1 scale based on acceptable and unacceptable performance, and combining them as a (weighted) average may be more practical for some users. Ultimately, users should choose a ranking system that is clear, justifiable and practical to implement for their application, recognising that greater methodological complexity does not necessarily yield better outcomes.

## 3.3 Steps

The importance of each step to climate services for agriculture, statistical methods used, and possible results are described, as well as how they should be interpreted and how they influence the analyses in subsequent steps.

### *3.3.1 Data selection and pre-processing*

This preliminary step involves making key decisions about the data to be used and the evaluation approach. This step involves: i) defining the geographic region of the evaluation, ii) determining the temporal scale suitable for the application, iii) determining which SRE data and station data to use within this region, and iv) determining the spatial scale for the evaluations (pixel-to-pixel or pixel-to-point).

The geographic region for the study should been defined by the intended use of the SRE data, with products ideally being evaluated over the area in which they will be used. However, in practice this choice may also be influenced by the availability of data with compromises sometimes needed.

The choice of temporal scale to carry out the evaluations should be determined by the needs of the application. For example, if the application requires calculation of summaries, such as start and end of the rainy season, or lengths of wet or dry spells, then data available at least daily will be required, whereas applications requiring only monthly totals may consider starting with aggregated pentadal, decadal, or monthly totals. The temporal scale also determines which rainfall summaries can be derived and, consequently, which evaluation metrics are appropriate. For example, metrics related to rainfall occurrence or spell persistence require daily data.

A literature review can determine the available SRE products for the defined geographic region. Key considerations for selection of SRE products should include reviewing the product's key properties, such as spatial and temporal resolution and length of available time series, to determine if they are compatible with the intended application. For example, for the calculation of average long term risks and climatology calculations, at least 30 years of data is generally recommended. The temporal resolution should be consistent with the requirements of the applications discussed above. Latency may also be a factor if the application requires recent data to be available within a specific time frame. Reading the technical documentation for SRE is also encouraged to understand known strengths and limitations and recommended uses, the methods used to generate the data, and whether gauge data are used as part of the method (e.g. by merging or for calibration).

Selection of station locations for gauge data should also be guided by temporal resolution and length to coincide with the SRE data and application needs, as well as obtaining a distribution of stations that represent the geographic area as well as possible (e.g. choosing stations in each climatic zone). If using SRE where gauge data are merged, then having some independent stations for comparison is desirable. Data quality also forms part of the selection process, which can vary between stations.

Determining data completeness is important (e.g. with an inventory analysis). This could lead to periods within a record or an entire record being discarded if the data are not sufficiently complete (e.g. at least 80%). Standard rainfall quality control checks can also help assess data quality and identify outliers to be removed or corrected (World Meteorological Organization, 2021).

The spatial scale for the evaluations also needs to be determined. Pixel-to-pixel comparisons requires gridding the gauge data to a common scale with the SRE pixels. This is often done through spatial interpolation methods, such as inverse distance weighting, nearest neighbour or kriging (Vicente-Serrano, et al., 2003; Maidment, et al., 2013). This typically requires a dense station network for accurately interpolating to the grid. The SRE data may also need to be rescaled to a common grid, often by simple pixel averaging. Pixel-to-point comparisons simply require extracting the SRE pixels which contain the gauge data locations to be compared directly with the gauge data.

The choice of comparison methods should depend on the specific application. Pixel-to-pixel comparisons are suitable when the SRE are to be used spatially over an area, such as for region-wide crop yield estimation, drought monitoring, or food security early warning systems. For these cases, it is important that the SRE represent area average rainfall characteristics and spatial variability at the grid scale, rather than rainfall at individual locations. Conversely, pixel-to-point comparisons are suitable when SRE are to be used to represent the climate at a point, such as for input to field-level crop models or farm-level climate services. For these applications, the evaluation seeks to determine whether the SRE data provide a representative estimate of local rainfall at a point, consistent with the way in which station data are currently used, and whether it can reliably be used in place of point data. The scale mismatch inherent in pixel-to-point comparisons should be acknowledged, with biases in rainfall occurrence and extremes expected between area and point measurements, which are discussed in subsequent steps of the methodology.

Regardless of the choice of comparison approach, the subsequent evaluation steps can utilise the same summaries and metrics for pixel-to-pixel and pixel-to-point data. It is typical in pixel-to-pixel comparisons to only select pixels containing at least one gauge, with evaluations done per pixel.

These preparatory decisions ensure that subsequent evaluation steps are based on appropriate, high-quality inputs and are properly aligned with the agricultural applications being assessed.

### *3.3.2 Spatial and temporal consistency checks*

This step assesses the spatial and temporal consistency of the SRE over the study area to ensure that there is consistency in localised information and does not involve comparison with station data.

For spatial consistency, long term averages of the annual total rainfall, number of rain days and mean rainfall per rain day are calculated for each SRE pixel. Mapping these summaries provides a preliminary assessment of how well the SRE captures climatology. This check can also highlight unnatural spatial variations, such as abrupt value changes between neighbouring pixels or over short distances without geographic explanation. Such inconsistencies may be due to the merging of multiple data sources in the SRE, especially if gauge data are used and users should likely not consider using SRE products that show clear spatial inconsistencies.

Temporal consistency checks focus on individual pixel SRE time series and aim to identify sudden step changes that cannot be accounted for. This is particularly relevant for reanalysis and merged satellite-gauge products that may be affected by changes in sensors, gauge inputs, or processing algorithms over time, and is a crucial evaluation step for applications requiring consistent, long time series.

Visual inspection through time series plots of annual or seasonal totals at pixel locations provides an initial indication of unrealistic long term trends or abrupt shifts. The temporal behaviour can be formally evaluated using well established statistical homogeneity and change-point tests widely applied in climate data analysis. The non-parametric Pettitt test is frequently used to detect single abrupt shifts (Pettitt, 1979), while the von Neumann ratio test identifies disruptions in the randomness structure of a series, (von Neumann, 1941). Parametric approaches, such as the Standard Normal Homogeneity Test (SNHT) (Alexandersson, 1986) and the Buishand tests (Buishand, 1982) are also commonly applied to detect shifts in the mean of climatic time series. These methods can be applied individually or combined to increase robustness and reduce the risk of false detections. For example, (Wijngaard, et al., 2003) applied a suite of homogeneity tests and classified each climatic series based on the level of agreement across methods, thereby strengthening confidence in detected breakpoints. Using multiple tests in combination, therefore, provides a more rigorous assessment of temporal stability and helps determine whether an SRE product is sufficiently consistent to proceed to the quantitative comparison steps that follow. Temporal inconsistencies, such as sudden jumps between years or changes in variability unrelated to climate, may indicate algorithm updates, sensor replacements, or changes in the underlying gauge network. Detecting these issues early helps determine whether the SRE is suitable for the detailed comparison steps that follow.

### 3.3.3 Quantitative SRE-gauge comparisons

These comparisons assess SRE performance against station data across key components that are of broad importance for agricultural applications. The comparisons involve systematically analysing annual summaries, seasonality, memory/spell persistence, rainfall amounts and extremes. Many application specific summaries depend on one or more of these components. Hence the performance at this step provides users with a general overview of the suitability of an SRE. Performance issues and biases identified at this stage suggest further investigation, including the potential need for bias correction methods to be considered in the next step.

Comparisons can be implemented using either pixel-to-pixel or pixel-to-point approaches depending on the choices and pre-processing done in the previous step. The procedures of evaluating the various components and metrics used remain the same irrespective of the comparison approach selected.

#### 3.3.3.1 Standard summaries

For agricultural applications, rainfall information is often linked to seasonal rainfall and cropping seasons. Many applications use rainfall information to determine “good” and “bad” agricultural seasons, making comparison of annual summaries crucial for evaluating SRE. Annual may refer to the calendar year, or a 12 month period adjusted to align with the main rainfall/cropping seasons.

Annual values of the standard summaries are calculated for the SRE and gauge data at each location. Percent bias, ratio of standard deviation and correlation are calculated from the yearly values to understand bias, variability and correlation. Time series graphs of the annual summaries are also produced to assess performance.

Good performance for all summaries and in all metrics would be highly encouraging. Where correlations are high, but bias is significant, the need for further evaluation might be necessary, including seasonality, and potential adjustments. This is commonly observed for number of rain days and mean rainfall per rain day due to differences between point and areal based measurements. Because bias correction does not improve correlation, low correlation with gauge data and large biases on these core summaries suggest that the SRE would not perform well for these summaries and might not be considered further.

#### 3.3.3.2 Seasonality

It is important for agricultural applications that SRE have accurate seasonality patterns because within season rainfall variation impact soil water content and crop growth (Zeppel, et al., 2014).

We propose fitting of models to investigate rainfall occurrence and amounts. Markov chain models have been widely used to model the sequence of rain days and dry days of rainfall gauge data (Stern & Coe, 1984; Deni, et al., 2009; Jimoh & Webster, 1996; da Silva Jale, et al., 2019) and can provide a detailed understanding of rainfall occurrence. However, their application in evaluating rainfall estimates remains limited. Acharya et al. (2019) used transition probabilities to evaluate rainfall estimates in Australia but without model fitting. The simplest of these models, often called a zero-order Markov chain model, estimates rainfall occurrence as a proportion for each day of the year, using logistic regression and Fourier series to model the periodic nature of the seasonality. The formula for a zero-order Markov chain model with three harmonics is shown in Equation (6).

$$r_i \sim \sum_{k=1}^{3}\left(\sin\left(\text{doy}.\,k.\frac{2\pi}{366}\right) + \cos\left(\text{doy}.\,k.\frac{2\pi}{366}\right)\right) \quad (6)$$

$$r_i = \begin{cases} 1 \text{ if rainfall on day i} \geq \text{threshold} \\ \quad 0 \text{ otherwise} \end{cases}$$

$\text{doy}$ = the day of the year from 1 to 366

Comparing these models for SRE with the gauge data allows for comparison of the occurrence of rain days throughout the year, showing how well SRE capture seasonal patterns. If seasonality is well captured, then biases can be studied and possible adjustments to the rain day threshold can be explored. This approach can be done by comparing models with increasing rain day thresholds for the SRE. Additionally, this approach helps to determine whether any bias varies seasonally, and with sufficient station coverage, whether it is dependent on location. Similar models can also be applied to compare rainfall amounts.

#### 3.3.3.3 Memory and spell persistence

Daily rainfall data often show serial correlation in rainfall occurrence and amounts (Stern & Coe, 1984; da Silva Jale, et al., 2019). This fact is an important aspect of agriculture-specific rainfall summaries. For example, the start of the rainy season is often defined in terms of a running rainfall total and a rain day condition, for example “the first occurrence after date d where total rainfall over y days exceeds x mm and where v out of w days were rain days” and can include an additional dry spell condition, for example “such that there is no dry spell of length x days in the following z days” (Stern & Cooper, 2011). Hence, understanding the persistence of wet and dry spells of SRE in comparison with gauge data through serial correlation will provide an indication of the performance of SRE in estimating these types of summaries.

A first-order Markov chain model of rainfall occurrence models the probability of rain given the state of the previous day. The formula for a first order Markov chain model with three harmonics is shown in Equation (7). This model fits two probability curves: 1) rain given rain - the probability of a rain day given the previous day was a rain day, indicating persistence of wet spells and 2) rain given dry – the probability of a rainy day given the previous day was a dry day, indicating the likelihood of a dry spell ending. By comparing first order Markov Chain models for gauge data and SRE, we can assess the performance of SRE in capturing seasonality patterns.

$$r_i \sim \mathrm{r}_{\mathrm{i}-1} * \sum_{k=1}^{3}\left(\sin\left(\text{doy}.\,k.\frac{2\pi}{366}\right) + \cos\left(\text{doy}.\,k.\frac{2\pi}{366}\right)\right) \quad (7)$$

$$r_i = \begin{cases} 1 \text{ if rainfall on day } i \geq \text{threshold} \\ \quad 0 \text{ otherwise} \end{cases}$$

$$r_{i-1} = \begin{cases} 1 \text{ if rainfall on day } i-1 \geq \text{threshold} \\ 0 \text{ otherwise} \end{cases}$$

doy = the day of the year from 1 to 366

As with the zero-order models, when the seasonal patterns are similar, we can observe the magnitude of any biases and investigate whether higher rain day thresholds for the SRE can correct this, and consider if the rain day threshold adjustments need to depend on seasonality, location, or state of the previous day. Hence a bias in rain day probabilities may not exclude a SRE from use when bias correction is possible.

In some locations, memory may persist beyond the previous day, suggesting that higher order Markov chain models could provide further insights into spell persistence and improve SRE evaluation.

#### 3.3.3.4 Rainfall amounts and extremes

Comparison of rainfall amounts assesses how well SRE detect different types of rainfall, ranging from dry days and light rain to violent rain and extreme rainfall events. The timing and distribution of rainfall amounts within the cropping season influence crop growth, in addition to the total rainfall over the period.

We assess detection across rainfall amount categories defined as: x < 0.85mm (dry), 0.85mm <= x < 5mm (light rain), 5mm <= x < 20mm (moderate rain), 20mm <= x < 40mm (heavy rain) and x >= 40mm (violent rain) where x denotes daily rainfall. These classifications are based on previous studies and adapted from WMO classifications (Acharya, et al., 2019; Zambrano-Bigiarini, et al., 2017) and can be adapted for local contexts.

Differences in distribution of rainfall amounts between gridded and gauge data are expected due to their differing spatial representations. In addition, comparisons at the daily time scale are also challenging for SRE to perform well on, due to the small margin of error compared with annual and monthly summaries, where aggregation increases skill. Understand these limitations is important, however, because this knowledge helps to quantify expected differences between rain gauge data and also supports appropriate use of SRE in applications that are sensitive to daily rainfall amounts.

SRE performance in extremes are also assessed using standardised indices, often derived on an annual basis. These include maximum daily rainfall, upper-percentile rainfall amounts, and counts of heavy rainfall days, as well as indicators of extreme dry conditions, such as the longest dry spell and the number of dry spells. Evaluating SRE performance using such indices complements the daily detection analysis by providing insight into how well SRE represent the behaviour of rainfall extremes relevant to agricultural decision making.

### 3.3.4 Bias correction of SRE

The potential need for adjustments to the SRE is discussed, based on the needs of the specific application and the results obtained in step 2.

Due to the areal nature of SRE, we expect higher rain day occurrence, particularly in pixel-to-point comparisons (see sections 4.3.1 and 4.3.2). This could affect key agricultural rainfall summaries, such as start and end of the rainy season and dry spell lengths.

If the general comparisons show that rain day occurrence from the SRE have high year-to-year correlation with the station data and capture well the seasonality of rain day occurrence, then bias correction methods could be applied to correct these differences.

Whether bias correction should be considered also depends on the specific application, even when clear biases have been detected. For example, for applications where absolute rainfall values are not used directly (e.g. anomalies from the climatology used for classification), then systematic biases may not affect this use. As there are risks that bias correction introduces other artifacts if implemented poorly, it should only be considered if necessary for the application.

Extensive literature exists on the use of bias correction methods of daily rainfall applied to SRE (Soo, et al., 2020; Ines & Hansen, 2006; Themeßl, et al., 2011; Piani, et al., 2010; Teutschbein & Seibert, 2012; Haerter, et al., 2011; Gumindoga, et al., 2019). Soo (2020) provides a comprehensive review of these methods. Some methods explicitly address biases in rain day occurrence, including various forms of distribution mapping (also known as quantile matching or probability mapping) (Ines & Hansen, 2006; Piani, et al., 2010; Teutschbein & Seibert, 2012) and Local amounts scaling method (LOCI) (Schmidli, et al., 2006).

Distribution mapping and LOCI are two step processes that start by calculating a rain day threshold for the SRE that matches a long-term gauge data rain day occurrence. It may be useful for the rain day threshold to be calibrated based on various factors. For example, Ines & Hansen (2006) calculate monthly rain day thresholds to account for seasonal differences. Seasonality comparisons can help to indicate what factors may need to be considered.

There is a risk of overfitting by making the calibration dependant on too many factors and therefore requiring many parameters (Schmidli, et al., 2006). Hence if the fitting remains reasonable, calibration based on less factors is desirable.

Location independence may be desirable to simplify the process, because location dependence over a large region could require more complex spatial analysis (e.g. interpolation of bias correction factors, although this has been done successfully) (Bhatti, et al., 2016; Gumindoga, et al., 2019).

Once thresholds have been determined, the values can be adjusted using various methods. In the distribution mapping method (Ines & Hansen (2006)) and the LOCI (Schmidli, Frei, & Vidale (2006)), amounts below the calibrated threshold are set to zero. Distribution mapping adjusts amounts above the threshold by transforming them from an empirical or gamma distribution of the estimates to a similar distribution of the gauge data. On the other hand, the LOCI method adjusts the amounts above the threshold by a scale factor calculated as the ratio of average estimates amounts (above the calibrated threshold) to the rain gauge amounts.

Soo (2020) describes the advantages and disadvantages of some common methods and classifies them as either mean based or distribution based. LOCI is mean based because it focuses on correcting long term mean rainfall amounts whereas the distribution mapping method in Ines & Hansen (2006) is distribution based as it aims to correct the distribution of rainfall amounts. Distribution based methods generally result in a rainfall distribution similar to the gauge data, including extremes, whereas mean based methods do not correct biases in extremes, due to their scaling transformation. Mean based methods, by definition, preserve the long term rainfall amounts of the gauge data.

Bias corrected SRE data can be assessed by repeating some of the quantitative comparisons. In particular, Markov chain models, mean rainfall amounts, and rain day occurrence comparisons will provide insights into how effectively biases have been corrected.

### 3.3.5 Application specific summaries

The final step considers comparisons of application specific summaries. Many agricultural applications also rely on tailored rainfall indices, highlighting the need to evaluate SRE performance on these application specific summaries. Results from the quantitative comparisons provide a general indication of SRE suitability. However, because SRE products are not validated for all possible agricultural applications, it is important to evaluate the SRE specifically on the summaries and events relevant to the intended context. Performance of general rainfall characteristics does not necessarily guarantee accurate estimation of application specific indicators, particularly when the definitions of these indicators are sensitive to small variations in rainfall values.

Common examples of such tailored summaries include the start, end and length of the rainy season, and the frequency and duration of dry spells. The precise definitions of these summaries can vary substantially across regions and user groups—for example, several definitions of the "start of the rains/season" may be used within a region and for different crops in the same region. In such cases, we recommend evaluating a representative set of definitions rather than repeating the full analysis if there are many small variations.

Table 4 provides examples of agricultural specific rainfall metrics that illustrate the breadth of possible applications. These include indices for season onset and cessation, dry spell characterisation, rainfall anomalies, drought monitoring and crop water availability. This non-exhaustive list highlights the diversity of rainfall based indicators that may be required for agricultural decision making and hence the importance of evaluating SRE performance for the specific summaries that matter for a wide range of agricultural applications.

Understanding how SRE perform under a range of metrics is important because it gives an indication of when SRE can be reliably used. For example, low correlation and large mean absolute error would indicate that the SRE may not be suitable for correctly detecting events in the same year. However, if the mean error is relatively small and the ratio of the standard deviations is close to 1, then this shows that the long term distribution is well represented. In such cases, SRE may still be suitable for communicating long term averages and risks of, for example early or late season onset, even if the year-to-year agreement with gauge data is low. This is particularly relevant for summaries, such as season onset or longest dry spell, which are highly sensitive to small differences in rainfall amounts and therefore difficult for SRE to have high correlation. These challenges should not rule out the use of SRE entirely, rather they emphasise the importance of evaluating suitability in relation to the specific application.

Table 4: Examples of agricultural specific rainfall metrics and their uses

| **Rainfall based Index** | **Typical Use in Agriculture** | **Reference** |
| --- | --- | --- |
| Start of rains / start of season<br>End of rains / end of season<br>Length of rainy season<br>Cumulative Rainfall Index (season total) | Planting decisions<br>Harvest planning; late-season risk assessment<br>Crop and variety suitability; cropping-calendar design | (Stern & Cooper, 2011) |
| Consecutive Dry Days Index (CDD)<br>Dry Spell Index (DSI) | Crop water stress<br>Crop insurance indicator | (Stern & Cooper, 2011), (Zhang, et al., 2011) |
| Effective Rainfall Index | Crop water availability assessments | (Allen, et al., 1998) |

| Palmer Drought Severity Index (PDSI) | Hydrological drought assessment; soil-moisture conditions | (Dai, et al., 2004) |
|---|---|---|
| Standardized Precipitation Index (SPI) | Seasonal drought monitoring | (World Meteorological Organization, 2012) |
| Crop-model yield estimates | Agronomic decision support | (Jones, et al., 2003) |
| Water-balance model outputs | Irrigation planning; soil-moisture deficits | (Allen, et al., 1998) |

# 4 Case studies

## 4.1 Data

### *4.1.1 Selected satellite and reanalysis rainfall estimates*

Five widely used SRE are selected to illustrate the methodology (TableTable 5). Three are satellite based estimates and two are reanalysis datasets (ERA5 and AgERA5). The criteria were SRE of daily values with few missing values, high spatial resolution, and a long record (preferably 30 years). All SRE have a spatial coverage that includes Africa, where most of the gauge data originate. The ERA5 hourly values were aggregated to daily values starting from 0600 UTC. NOAA RFE2, IMERG and ARC2 do not meet all of the criteria above but were considered and included in a selection of the analysis to highlight interesting results.

Table 5: Overview of five satellite and reanalysis rainfall datasets used to illustrate the methodology

| SRE | Method type | Coverage | Time period | Spatial resolution | Temporal resolution | Reference |
|---|---|---|---|---|---|---|
| CHIRPS v2.0 | Satellite + gauge merge | Global | 1983 – Present | 0.05° | Daily | Funk, et al. (2015) |
| ERA5 | Reanalysis | Global | 1940 – Present | 0.25° | Hourly | Hersbach, et al. (2020) |
| AgERA5 | Reanalysis | Global | 1979 – Present | 0.1° | Daily | Boogaard, et al. (2020) |
| ENACTS | Satellite + gauge merge | Selected countries | Various (often 1983 – near present) | 0.0375° | Daily | Dinku, et al. (2017) |

| TAMSAT v3.1 | Satellite + gauge calibration | Africa | 1983 – Present | 0.0375° | Daily | Maidment R. I., et al. (2017) |
|---|---|---|---|---|---|---|

### 4.1.2 Station data

Data from 12 stations in Africa and the Caribbean were used for comparison with the SRE (Table 6Table). The data were provided by Zambia Meteorological Department (7 stations), the Caribbean Institute for Meteorology and Hydrology (1 station), the International Crops Research Institute for the Semi-Arid Tropics (1 station) and Ghana Meteorological Agency (3 stations). The data consist of daily rain gauge measurements with records starting from 1983 to align with some SRE, although some stations' records started considerably earlier. The length of the records varies, depending on data availability, with the shortest ending in 2011 and the longest in 2022. The data were quality controlled before analysis, with any values failing quality control checks replaced by missing values. Most missing values resulted from gaps in records provided by the data sources.

This set of stations from diverse locations was selected to assess the appropriateness of the proposed methodology under varying climatic and geographic conditions and to illustrate the range of results that can be obtained, rather than being comprehensive at one particular location, as it may be used in practice.

Some of these data were used in generating some of the SRE. Nine of the 12 stations contributed to the CHIRPS product (https://data.chc.ucsb.edu/products/CHIRPS-2.0/diagnostics/list_of_stations_used/monthly/), but this does not include the full data records and the contribution of station data from Africa to CHIRPS has declined over time (Funk, et al., 2015). Additionally, CHIRPS merges station data on a 5 day basis (Funk, et al., 2015), meaning that many analyses of the daily values can be considered reasonably independent. Some of these data may have contributed to other SRE, but less directly; for example, TAMSAT includes station data for calibration purposes. Any dependencies should be recognised when evaluating and comparing SRE.

Table 6: Details of the 12 stations used to illustrate the methodology

| Country | Station | Latitude | Longitude | Period | Complete Days (%) |
|---|---|---|---|---|---|
| Barbados | Husbands | 13.15 | -59.62 | 1983 - 2019 | 100 |
| Ghana | Saltpond | 5.20 | -1.07 | 1983 - 2019 | 91.7 |
| Ghana | Tamale | 9.55 | -0.86 | 1983 - 2022 | 97.7 |
| Ghana | Wa | 10.05 | -2.50 | 1983 - 2020 | 90.4 |
| Niger | Sadore | 13.23 | 2.28 | 1983 - 2015 | 99.0 |
| Zambia | Choma | -16.84 | 27.07 | 1983 - 2011 | 99.7 |
| Zambia | Kasama | -10.22 | 31.14 | 1983 - 2018 | 95.7 |
| Zambia | Magoye | -16.00 | 27.62 | 1983 - 2016 | 97.7 |
| Zambia | Mansa | -11.14 | 28.87 | 1983 - 2017 | 91.3 |
| Zambia | Moorings | -16.19 | 27.53 | 1983 - 2010 | 100 |
| Zambia | Mpika | -11.90 | 31.43 | 1983 - 2020 | 90.3 |
| Zambia | Livingstone | -17.82 | 25.82 | 1983 - 2018 | 94.8 |

## 4.2 Statistical software

Summary statistics, calculations, graphs, and maps were generated using R 4.0.4 statistical software (R Core Team, 2021) with RStudio 1.4.1717 (RStudio Team, 2020) and the following packages: tidyverse (Wickham, et al., 2019), here (Müller, 2020), reshape2 (Wickham, 2007), viridis (Garnier, et al., 2021), hydroGOF (Zambrano-Bigiarini, 2020), knitr (Xie, 2018), kableExtra (Zhu, 2021), rnaturalearth, rnaturalearthdata (South, 2017), ggrepel (Slowikowski, 2021) and sp (Bivand, et al., 2013).

## 4.3 Results

### *4.3.1 Data selection and pre-processing*

The data choices outlined in Sections 4.1.1 and 4.1.2 reflect the decisions made as part of Step 0 of the methodology and are tailored to the requirements of the PICSA application. PICSA relies on long term, continuous daily rainfall station records. The selected SRE products provide daily data with sufficiently long temporal coverage for PICSA. In practice, this analysis would be done within a single specific region. The diversity of stations and SRE products selected allows the methodology to be illustrated across a range of climatic and geographic contexts. As the PICSA approach uses rainfall information for farm-level climate services, and the rainfall indices and summaries used in PICSA has been developed around point-based gauge data, pixel-to-point comparison has been selected as the appropriate evaluation approach.

### *4.3.2 Spatial consistency checks*

Unless otherwise stated, throughout these case studies, a rain day is defined as a day receiving at least 0.85mm of rainfall. This threshold is essentially equivalent to the commonly used 1mm threshold but helps reduce potential bias from round errors in station records (Stern & Cooper, 2011).

The maps of Zambia in Figure 1 and Figure 2 show the average annual total rainfall and average number of rain days respectively, from ENACTS (2019 version) (a), ENACTS (current) (b), and TAMSAT (c). Figure 3 compares the average rain per rain day between ENACTS (2019 version) (a), ENACTS (current) (b), and CHIRPS (c). All SRE appear to accurately represent Zambia's rainfall climatology, with high annual rainfall in the north of the country. The old ENACTS (ENACTS 2019) data seem to show unnatural spatial inconsistencies in all three rainfall summaries. We suspect that the circular anomalies, observed in similar locations in both maps, coincide with station locations where gauge data have been merged with satellite data without sufficiently accounting for differences between point-based station values and satellite pixel values. A similar problem was observed with the African Rainfall Climatology Version 2 (Novella & Thiaw, 2013) (not shown in this study) which also merges station and satellite data. It is reassuring that this issue does not appear present in the latest ENACTS data. No such spatial inconsistencies are observed for TAMSAT (Figures 1 (c) and 2 (c)) and CHIRPS (Figure 3 (c)), indicating that their station data calibration/merging algorithms account for differences between station and satellite data.

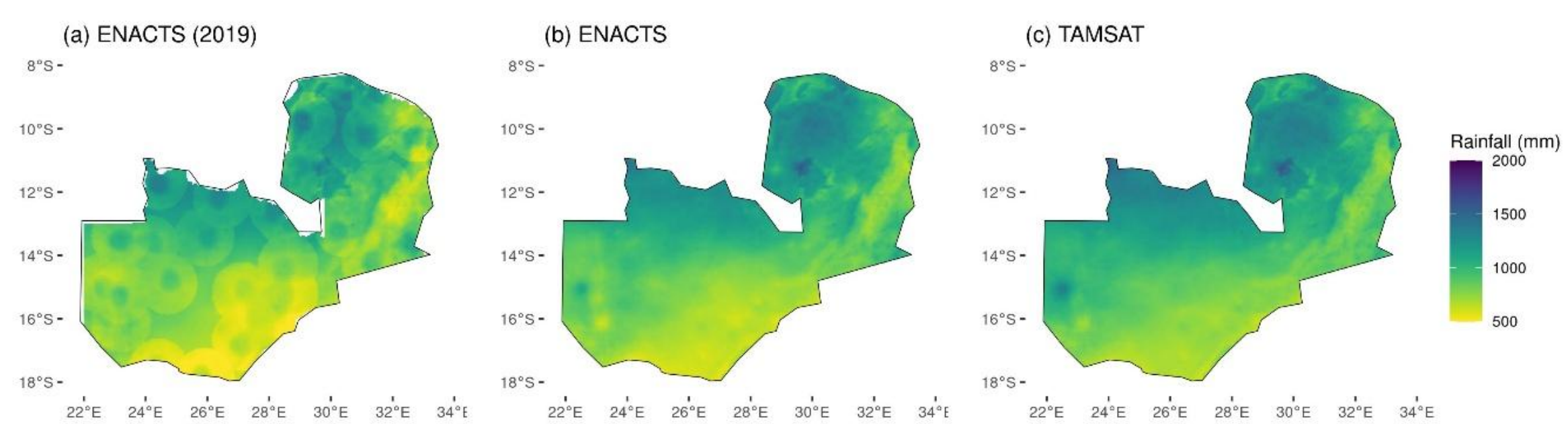

Figure 1: Mean annual total rainfall over Zambia from ENACTS (2019 version), ENACTS (latest version), and TAMSAT.

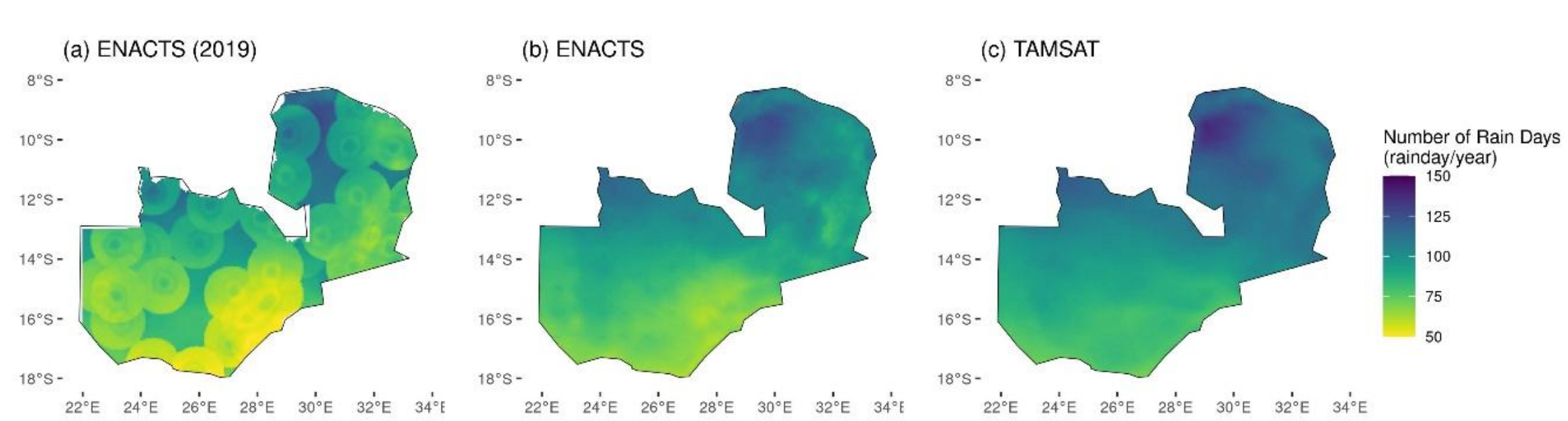

Figure 2: Mean annual number of rain days  over Zambia from ENACTS (2019 version), ENACTS (latest version), and TAMSAT.

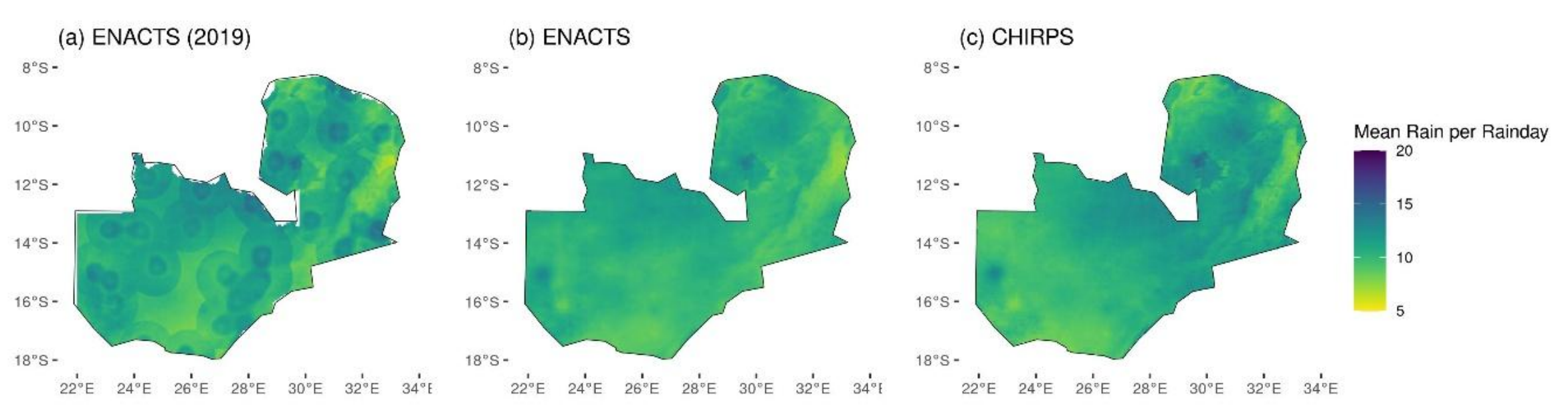

Figure 3: Mean rain per rainy day  over Zambia from ENACTS (2019 version), ENACTS (latest version), and CHIRPS.

### *4.3.3 Quantitative SRE-Gauge comparisons*

#### *4.3.3.1 Standard summaries*

Annual total rainfall (August – July) was calculated in years with at least 355 daily values. Comparison metrics are presented in Table 7.

Figure 4 shows the time series graphs of annual total rainfall from gauge, TAMSAT, and ERA5 when all sources had data. The horizontal lines indicate the mean annual rainfall over the same years for each source. The TAMSAT means are similar to the gauge data in all locations, with a small bias in some locations (pbias <= 7%) and good correlations (0.47 <= r <= 0.80). Kasama and Mansa have the lowest correlation, but the pattern of peaks and dips is generally in the same years across the stations. TAMSAT shows slightly less variability in all locations (0.64 <= rSD <= 0.84). While ERA5 also correlates well with the gauge data (0.42 <= r <= 0.82), it generally overestimates rainfall at all locations except Kasama.

In general, all estimates have a good positive correlation with the gauge data (Table 7). AGERA5, like ERA5, overestimates rainfall totals, but all other estimates have a low percentage bias, with a small underestimation in most places. Performance is worse at the northern stations with lower KGE values compared with the southern stations.

ENACTS and CHIRPS generally outperform the other estimates across most metrics. However, both ENACTS and CHIRPS include merged data from official stations, likely including 6 of the evaluated stations, which will account for some of their performance. But CHIRPS outperforms TAMSAT and ERA5 at Moorings station which is not part of the merging process.

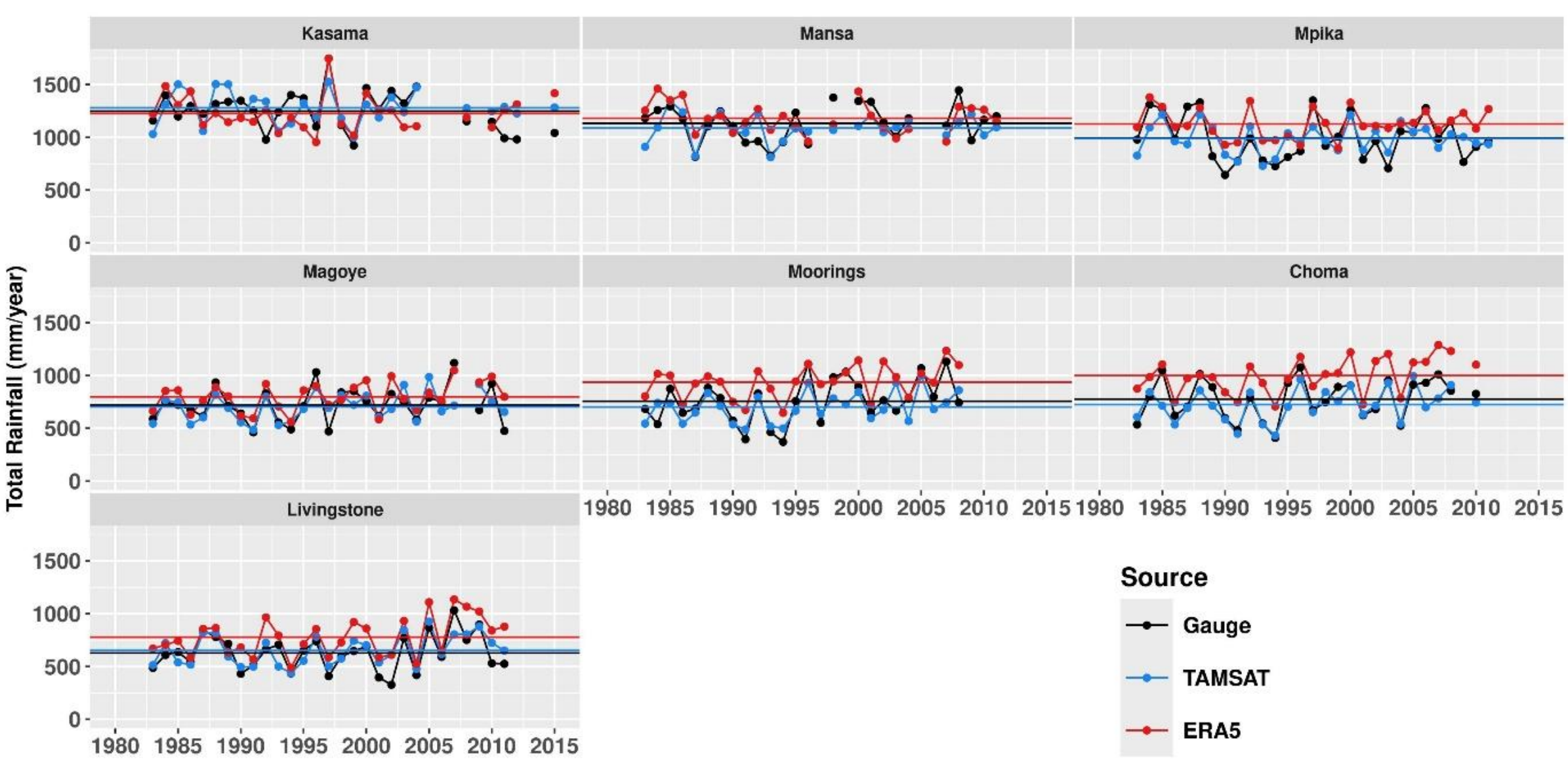

Figure 4: Total annual (August to July) rainfall from Gauge, TAMSAT and ERA5 at seven locations in Zambia

Table 7: Comparison metrics for total annual rainfall at seven locations in Zambia. Metrics shown are the Pearson correlation coefficient (r, unitless), percent bias (pbias, %), ratio of standard deviations (rSD, unitless), and modified Kling–Gupta efficiency (KGE, unitless)

| Metric | Station | CHIRPS | ERA5 | AGERA5 | TAMSAT | ENACTS |
|---|---|---|---|---|---|---|
| r | Kasama | 0.79 | 0.42 | 0.43 | 0.47 | 0.81 |
| r | Mansa | 0.76 | 0.48 | 0.44 | 0.5 | 0.84 |

| r | Mpika | 0.59 | 0.69 | 0.68 | 0.69 | 0.89 |
|---|---|---|---|---|---|---|
| r | Magoye | 0.87 | 0.76 | 0.74 | 0.59 | 0.87 |
| r | Moorings | 0.84 | 0.72 | 0.7 | 0.7 | 0.8 |
| r | Choma | 0.9 | 0.8 | 0.77 | 0.8 | 0.93 |
| r | Livingstone | 0.9 | 0.82 | 0.81 | 0.78 | 0.97 |
| pbias | Kasama | 4 | -2 | -2 | 3 | -4 |
| pbias | Mansa | -5 | 4 | 8 | -4 | -9 |
| pbias | Mpika | -2 | 14 | 26 | 0 | -15 |
| pbias | Magoye | -2 | 11 | 15 | -2 | -17 |
| pbias | Moorings | -5 | 24 | 30 | -7 | -20 |
| pbias | Choma | -4 | 29 | 22 | -7 | -18 |
| pbias | Livingstone | 4 | 24 | 30 | 4 | -12 |
| rSD | Kasama | 0.86 | 0.9 | 0.9 | 0.81 | 0.74 |
| rSD | Mansa | 0.63 | 0.83 | 0.9 | 0.74 | 0.79 |
| rSD | Mpika | 0.79 | 0.64 | 0.74 | 0.64 | 0.64 |
| rSD | Magoye | 0.83 | 0.81 | 0.88 | 0.8 | 0.99 |
| rSD | Moorings | 0.72 | 0.74 | 0.8 | 0.69 | 0.86 |
| rSD | Choma | 0.9 | 0.89 | 0.87 | 0.82 | 0.89 |
| rSD | Livingstone | 0.87 | 1.06 | 1.1 | 0.84 | 0.86 |
| KGE | Kasama | 0.75 | 0.41 | 0.42 | 0.43 | 0.68 |
| KGE | Mansa | 0.56 | 0.45 | 0.43 | 0.44 | 0.72 |
| KGE | Mpika | 0.54 | 0.51 | 0.52 | 0.52 | 0.59 |
| KGE | Magoye | 0.78 | 0.67 | 0.67 | 0.55 | 0.78 |
| KGE | Moorings | 0.67 | 0.55 | 0.53 | 0.56 | 0.68 |
| KGE | Choma | 0.85 | 0.64 | 0.66 | 0.72 | 0.77 |
| KGE | Livingstone | 0.83 | 0.69 | 0.64 | 0.72 | 0.81 |

A contrasting pattern of performance is observed at Husbands, Barbados. Figure 5 shows the comparisons of the annual total rainfall at Husbands, Barbados for two SRE, and corresponding metrics in Table 8. Both SRE show strong positive correlation with gauge data ($r >= 0.79$). But there is a large underestimation by the SRE, up to 30% for CHIRPS. This substantial bias, despite high correlation warrants further investigation.

To explore this discrepancy, Figure 6 compares the monthly totals for CHIRPS against gauge data. A large amount of the annual underestimation is attributed to the failure of CHIRPS to capture the most extreme monthly rainfall totals of more than 300 mm, primarily observed in October and November but not exclusively. This suggests that CHIRPS captures the general rainfall patterns throughout the year but may fail to detect the high intensity rainfall events, resulting in consistent underestimation.

An analysis of the daily rainfall events could be performed to understand SRE performance at estimating daily rainfall totals of different intensities. This analysis could be complemented by a comparison with known extreme rainfall events at this location.

This example underscores the importance of location specific evaluations. The performance of rainfall estimates can vary significantly, depending on a number of factors, including climate and terrain. Hence we suggest that SRE should always be evaluated at each intended application location rather than assuming consistent performance across all locations.

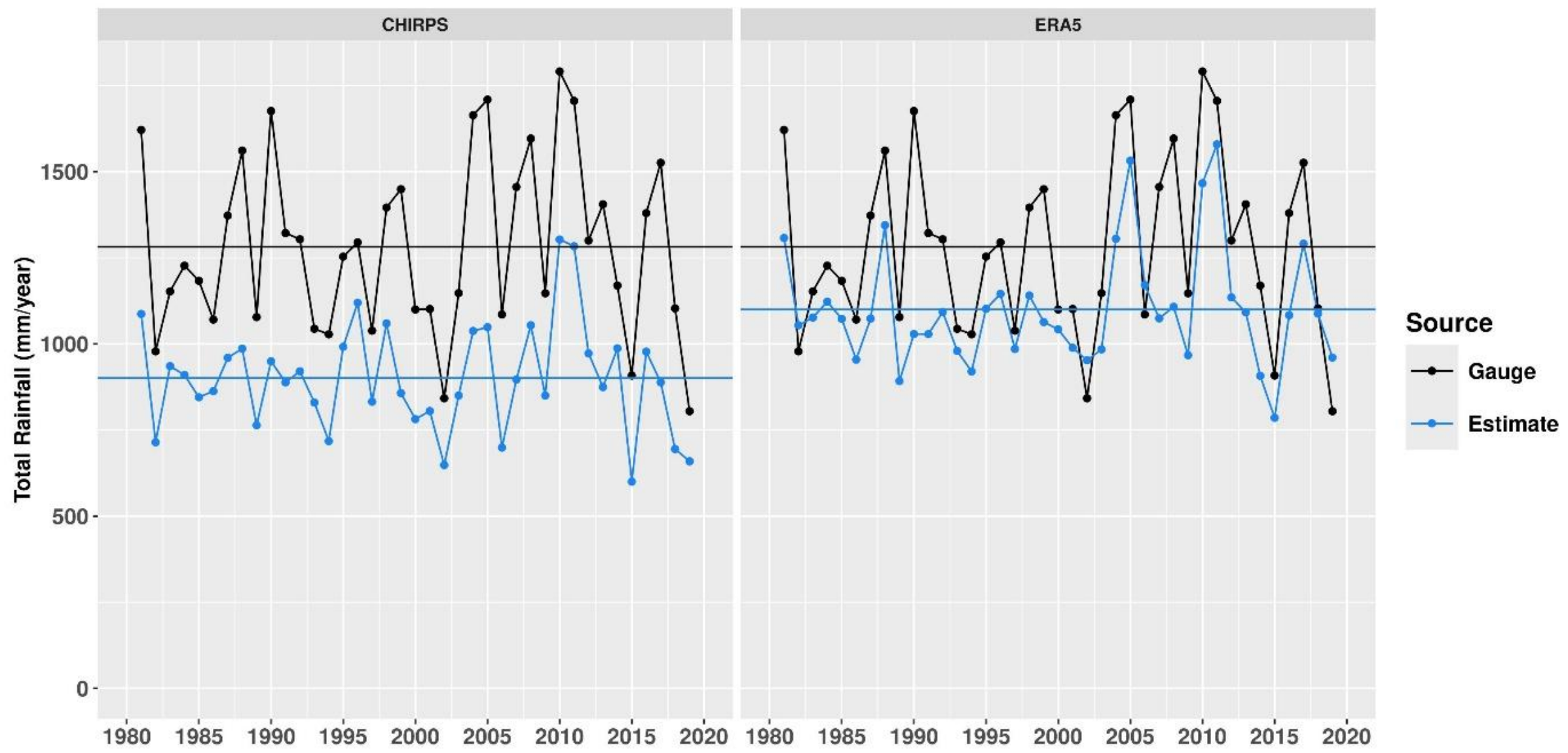

Figure 5: Total annual rainfall comparing Gauge data with CHIRPS and ERA5 at Husbands, Barbados

Table 8: Comparison Performance metrics for total annual rainfall at Husbands, Barbados. Metrics shown are the correlation coefficient (r, unitless), percent bias (pbias, %) and mean absolute error (MAE, mm).

| Metric | CHIRPS | ERA5 |
|---|---|---|
| r | 0.83 | 0.79 |
| pbias | -30 | -14 |
| MAE | 381 | 204 |

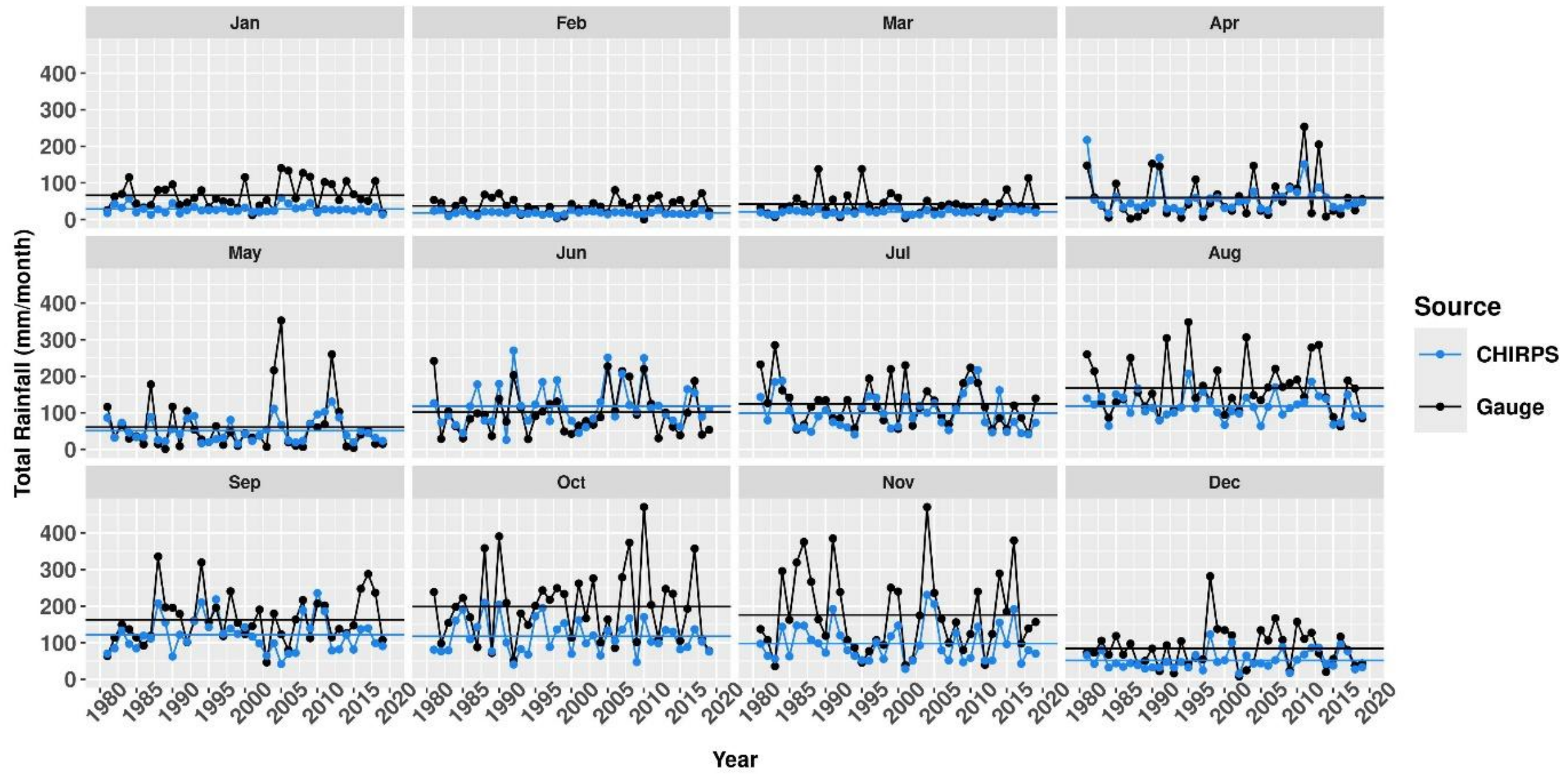

Figure 6: Total monthly rainfall comparing Gauge data with CHIRPS at Husbands, Barbados

Figure 7 shows annual number of rain days at Livingstone in comparison with five SRE. Metrics are provided in Table 9. All SRE overestimate the number of rain days, with mean absolute error ranging from 13 to 67 days per year. ERA5 estimates, which represent the largest pixel size, have the largest percentage bias (84%). The new ENACTS has the lowest percentage bias (8%), followed by CHIRPS (29%). All SRE also have a good positive correlation with the gauge data (0.63 <= r <= 0.83), with ERA5 having the lowest. There is a corresponding underestimation in mean rainfall amounts as shown in Figure 8, across the estimates. Again, there is a good positive correlation between the gauge data and estimates.

The results from Zambia show that annual rainfall totals broadly compare well across the stations for each SRE. While the rain day occurrence tends to be overestimated and mean rain per rain day underestimated, the correlation between gauge and estimates remains strong. This indicates that adjusting the rain day threshold in the estimates could improve the estimation of both rain day occurrence and mean rainfall per rain day.

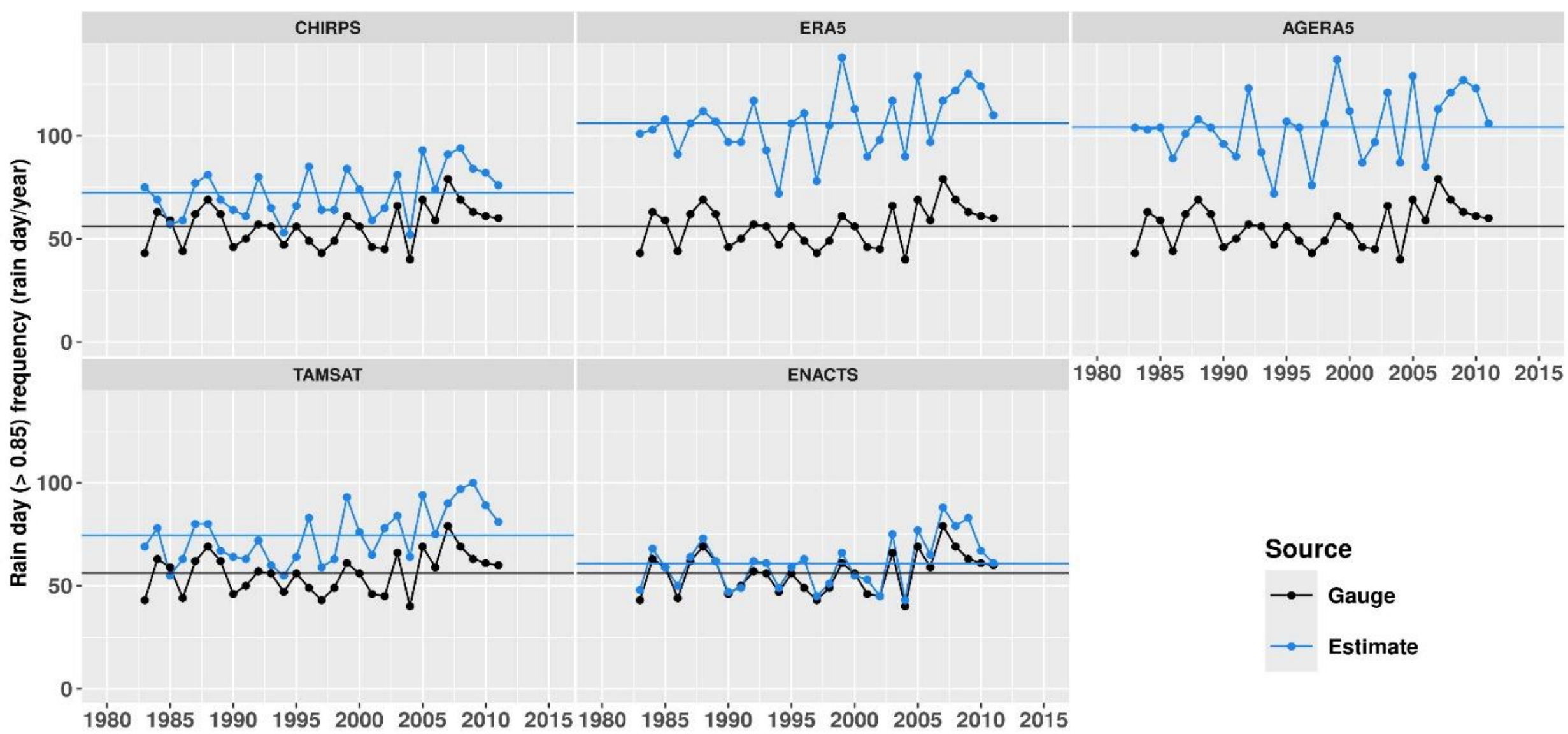

Figure 7: Annual (August to July) number of rain days comparing gauge data with estimates at Livingstone, Zambia

Table 9: Comparison metrics for number of rain days at Livingstone, Zambia. Correlation coefficient (r) is unitless, percent bias (pbias) is expressed as a percentage (%) and mean absolute error (MAE) is expressed in days.

| Metric | CHIRPS | ERA5 | AGERA5 | TAMSAT | ENACTS |
|---|---|---|---|---|---|
| r | 0.74 | 0.7 | 0.65 | 0.67 | 0.93 |
| pbias | 29 | 89 | 86 | 33 | 8 |
| MAE | 16 | 50 | 48 | 19 | 5 |

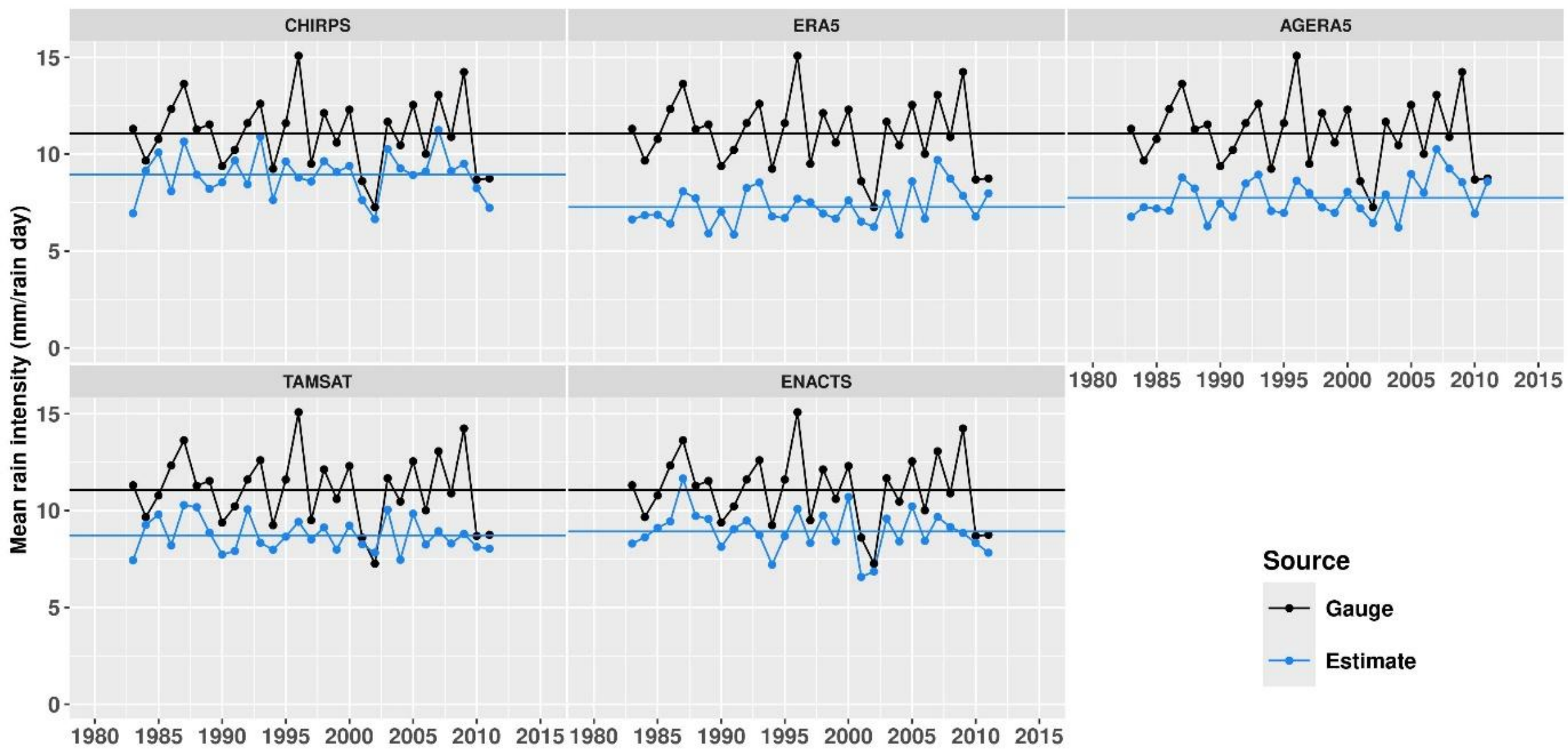

Figure 8: Annual mean rainfall intensity comparing gauge data with estimates at Livingstone, Zambia

#### *4.3.3.2 Seasonality*

Zero order Markov chain models with three harmonics were fitted separately to the rainfall occurrence at each of the seven stations in Zambia for the gauge data and each SRE, with 1 August considered as the first day of the year. At each location only days with non-missing values for both the gauge and SRE values were included in the model to ensure comparison on the same days. Figures 9–11 show the fitted rain day occurrence for the gauge data using a 0.85mm rain day threshold (solid black line) and for TAMSAT, CHIRPS, and ERA5 respectively, with various increasing rain day thresholds (coloured dashed lines) ranging from 0.85mm to 5mm.

The shape of the curves indicates that seasonality is well captured by each of the SRE, with the highest rain day occurrence in the November to May rainy season. At the 0.85mm threshold, all SRE consistently overestimate rain day occurrence, with ERA5 showing the largest overestimation and CHIRPS and TAMSAT the lowest, consistent with the findings in section 3.2.1. At higher thresholds the SRE match the gauge data more closely while maintaining the same general shape. For example, ERA5 with a 5mm threshold closely matches the gauge data across all stations although it still slightly overestimates the peak of the rainy season at Moorings and Mpika.

For some SRE, the graphs show that the optimal threshold to match the gauge data may vary by location. For example, at Choma, a threshold between 2 mm and 3 mm appears optimal for TAMSAT, whereas at Moorings, a threshold between 4 mm and 5 mm provides a better match. The optimal threshold may also vary by time of year. At Choma, both CHIRPS and TAMSAT overestimate rain day occurrence in the early part of the rainy season (November–December) even at a 5 mm threshold, but slightly underestimate it in the late season (April–May) even with a 0.85mm threshold.

Figure 12 presents a similar analysis for four stations in West Africa, comparing CHIRPS data using different thresholds. At Wa, CHIRPS estimates the number of rain days well in the rainiest months (August-October) but overestimates in the beginning of the rainy season (April-July), especially at lower thresholds. At the coastal location of Saltpond, the biases show a different structure, with CHIRPS overestimating rainfall occurrence in the first half of the year and underestimating it between the two rainy seasons.

For ERA5 at the same stations (not shown), there is a major overestimation at the 0.85mm threshold, particularly at Saltpond. In June, for instance, ERA5 estimates a rain day occurrence of 0.95 at the 0.85mm threshold and 0.7 at the 2mm threshold, compared with 0.4 from the gauge data at the same 0.85mm threshold. Caution is needed when analysing SRE at coastal locations where the nearest gridded data pixel may be influenced by ocean coverage, resulting in a distorted rainfall pattern compared with gauge data. In such cases, pixels located slightly inland may provide a more representative match to gauge data.

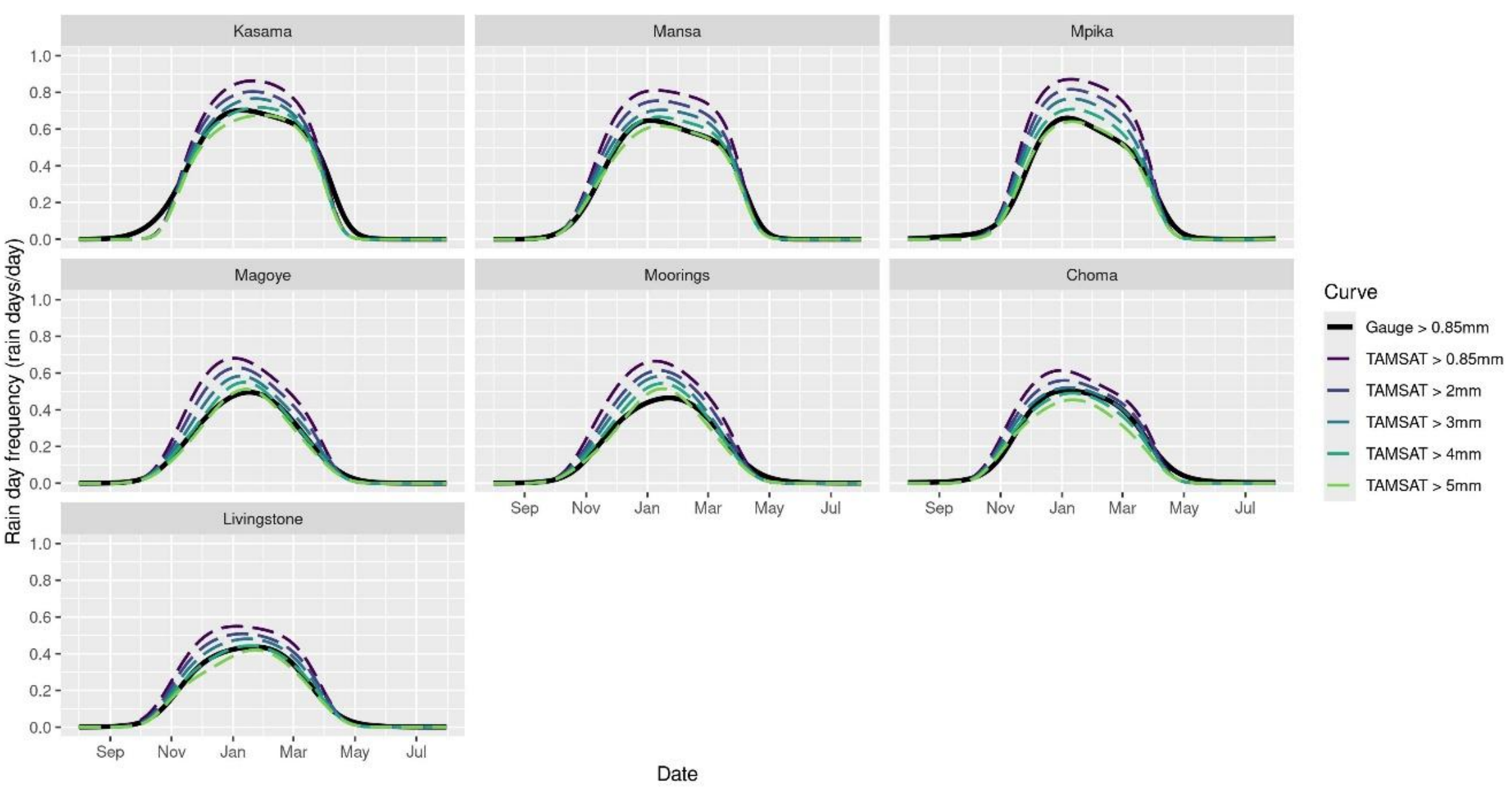

Figure 9: Gauge and TAMSAT estimated rain day frequency (rain days/day) at various thresholds at seven locations in Zambia

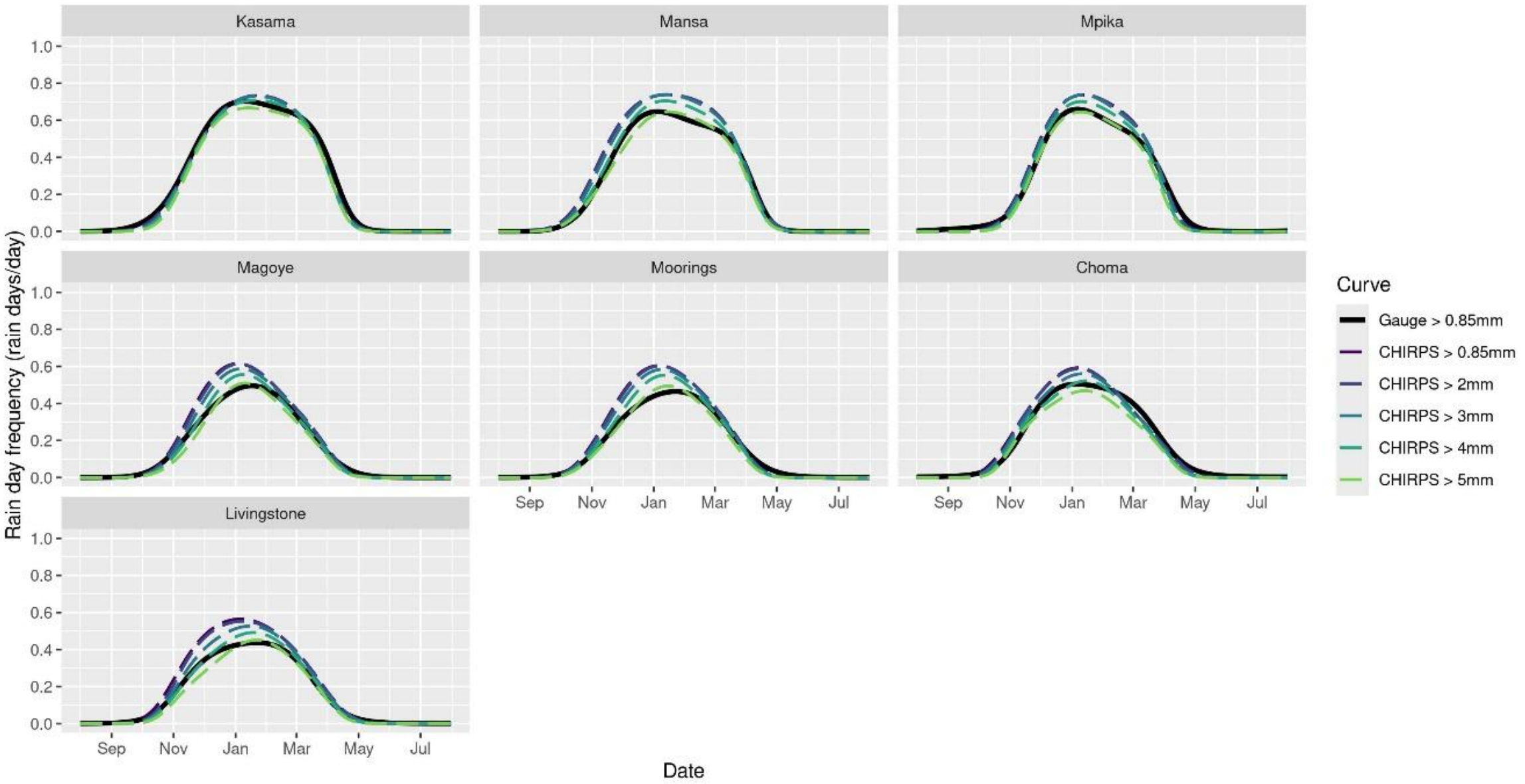

Figure 10: Gauge and CHIRPS estimated rain day frequency (rain days/day) at various thresholds at seven locations in Zambia

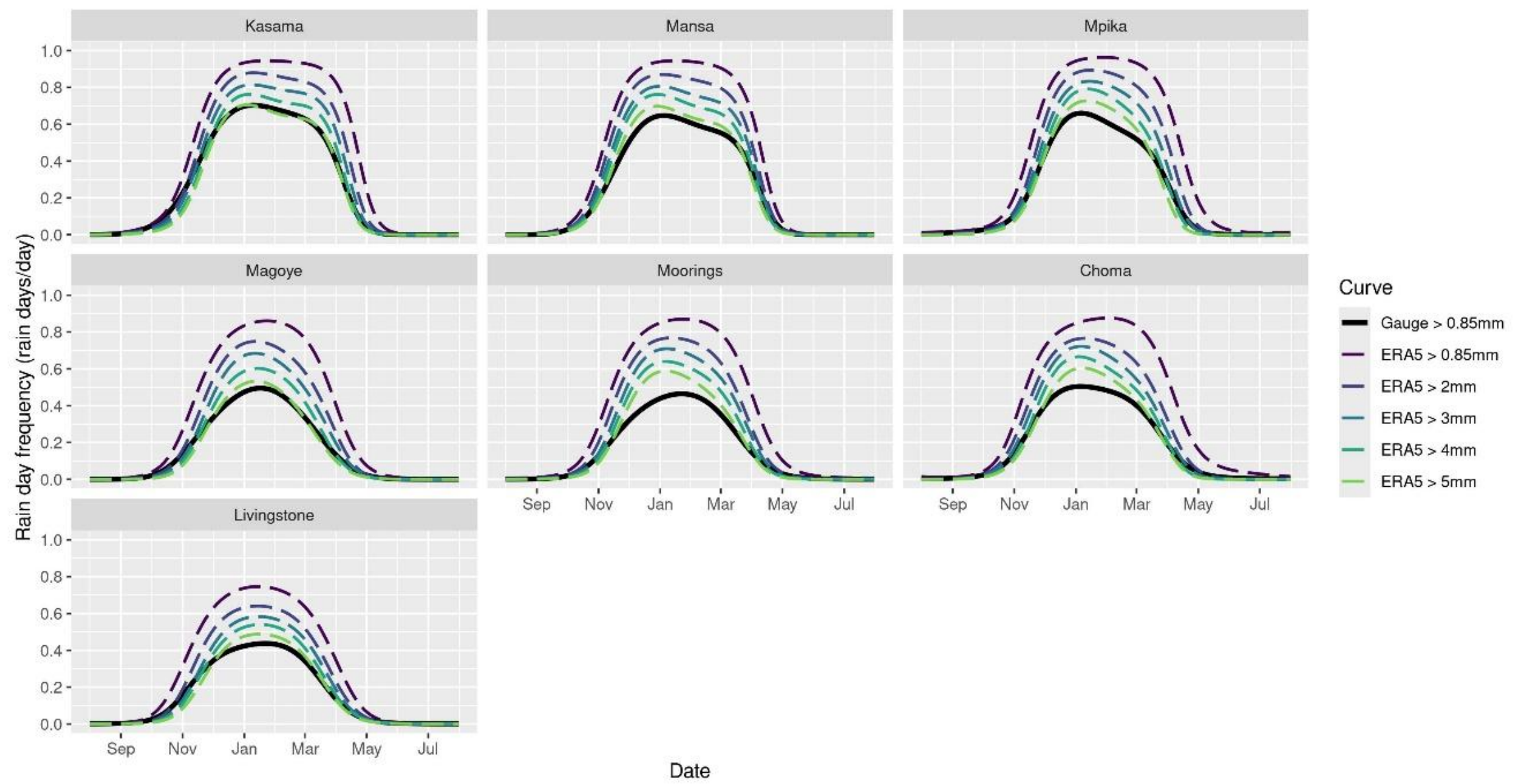

Figure 11: Gauge and ERA5 estimated rain day frequency (rain days/day) at various thresholds at seven locations in Zambia

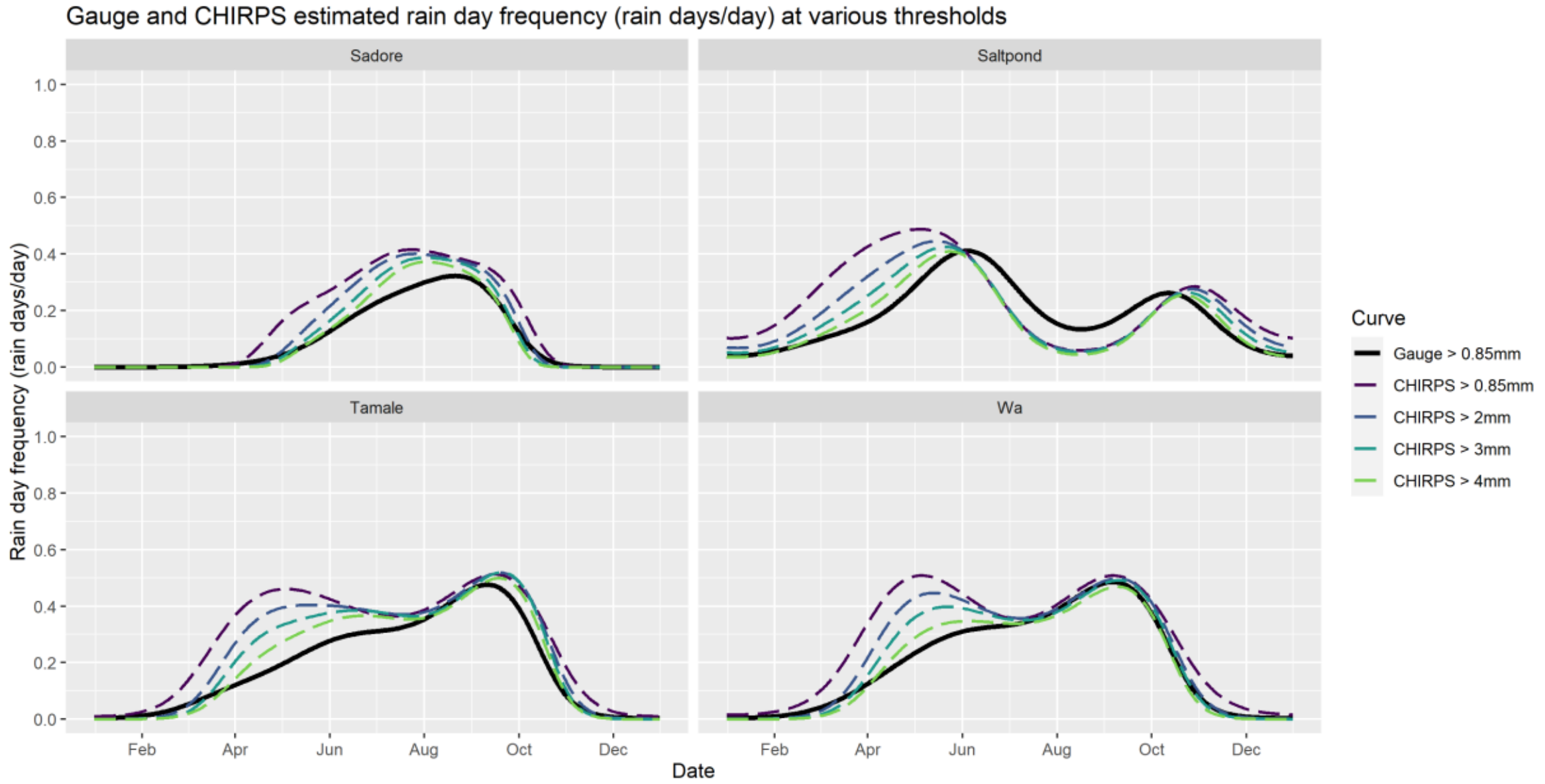

Figure 12: Gauge against CHIRPS estimated rain day frequency (rain days/day) at various thresholds at four locations in West Africa

#### *4.3.3.3 Memory and spell persistence*

First-order Markov chain models with three harmonics were fitted separately to the rainfall occurrence at each of the seven stations in Zambia for both the gauge data and each SRE, with 1 August defined as the first day of the year. At each location only days with non-missing values in both gauge and SRE data were included in the model to ensure comparison were made on the same set of days. Separate models were also fitted for the SRE using different rain day thresholds of 0.85mm, 2mm, 3mm, 4mm and 5mm.

All SRE show a consistent seasonality pattern using the same rain day threshold, as shown for CHIRPS and TAMSAT in Figures 13 and 14, respectively. As with the zero order models, SRE tend to overestimate rainfall occurrence during the peak rainy months. This overestimation is particularly evident in the rain given rain probabilities.

CHIRPS gives similar rain given dry probabilities to the gauge data across most stations, with some overestimation. TAMSAT also show similar patterns at some stations. This translates into a higher probabilities of wet spells continuing (rain given rain) in the SRE but similar probabilities of dry spells ending (rain given dry) comparable with the gauge data. At higher rain day thresholds for the SRE data, the probabilities in both sets of curves are reduced, thereby decreasing the high bias in the rain given rain probabilities. This potentially indicates that adjusting the rain day threshold can improve the estimation of both transition probabilities for some SRE. For example TAMSAT with a 4mm rain day threshold, shown in Figure 15 provides improved estimates of both states. However, for ERA5 (not shown), the results suggest that different rain day thresholds may be needed to improve the estimation of the two transition probabilities.

The figures display probabilities only for the rainy season (November to May) because the limited number of rain days outside of this period could result in unstable estimates.

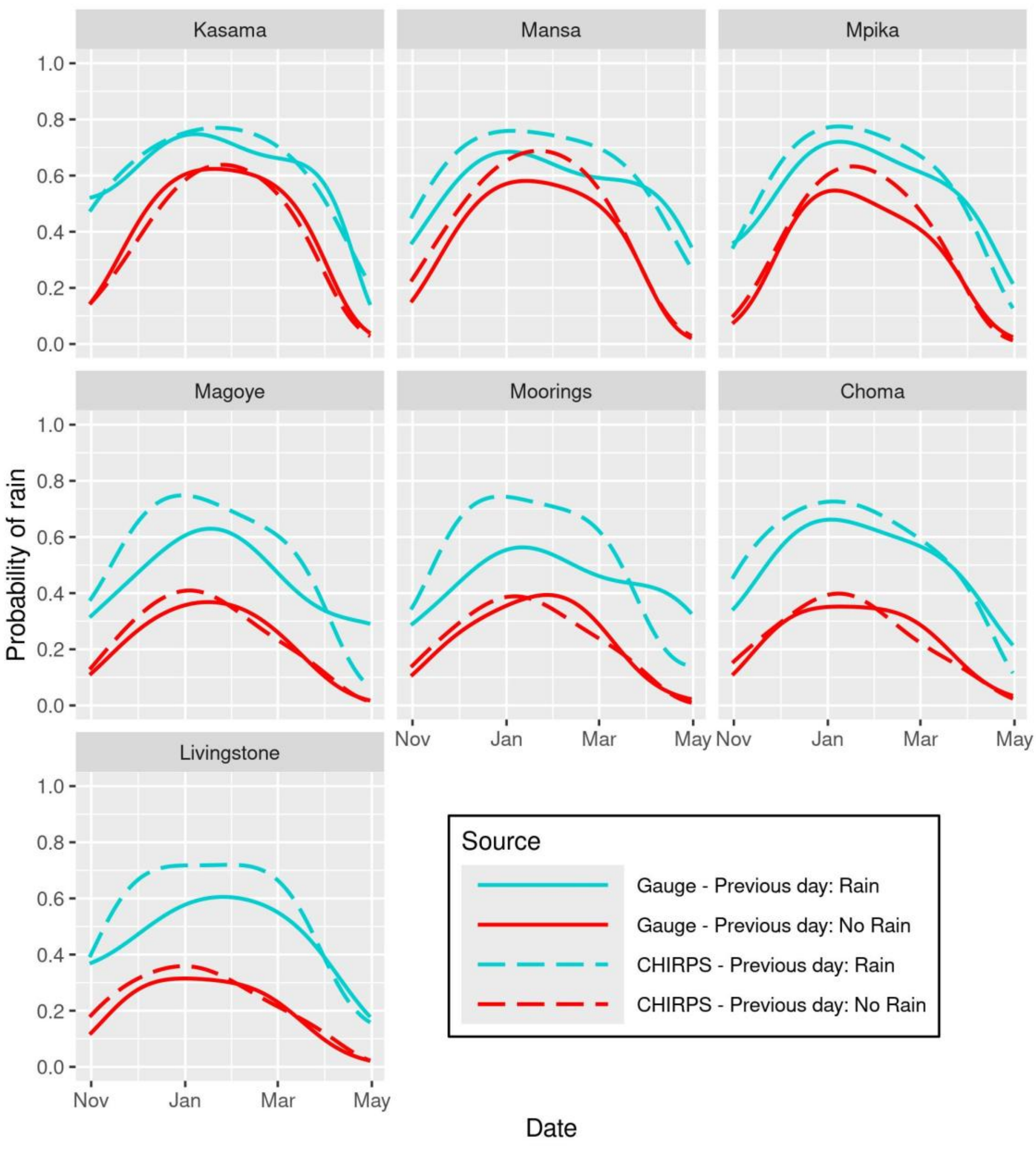

Figure 13: The probability of rain given the previous day being dry or wet for Gauge vs Chirps at seven locations in Zambia

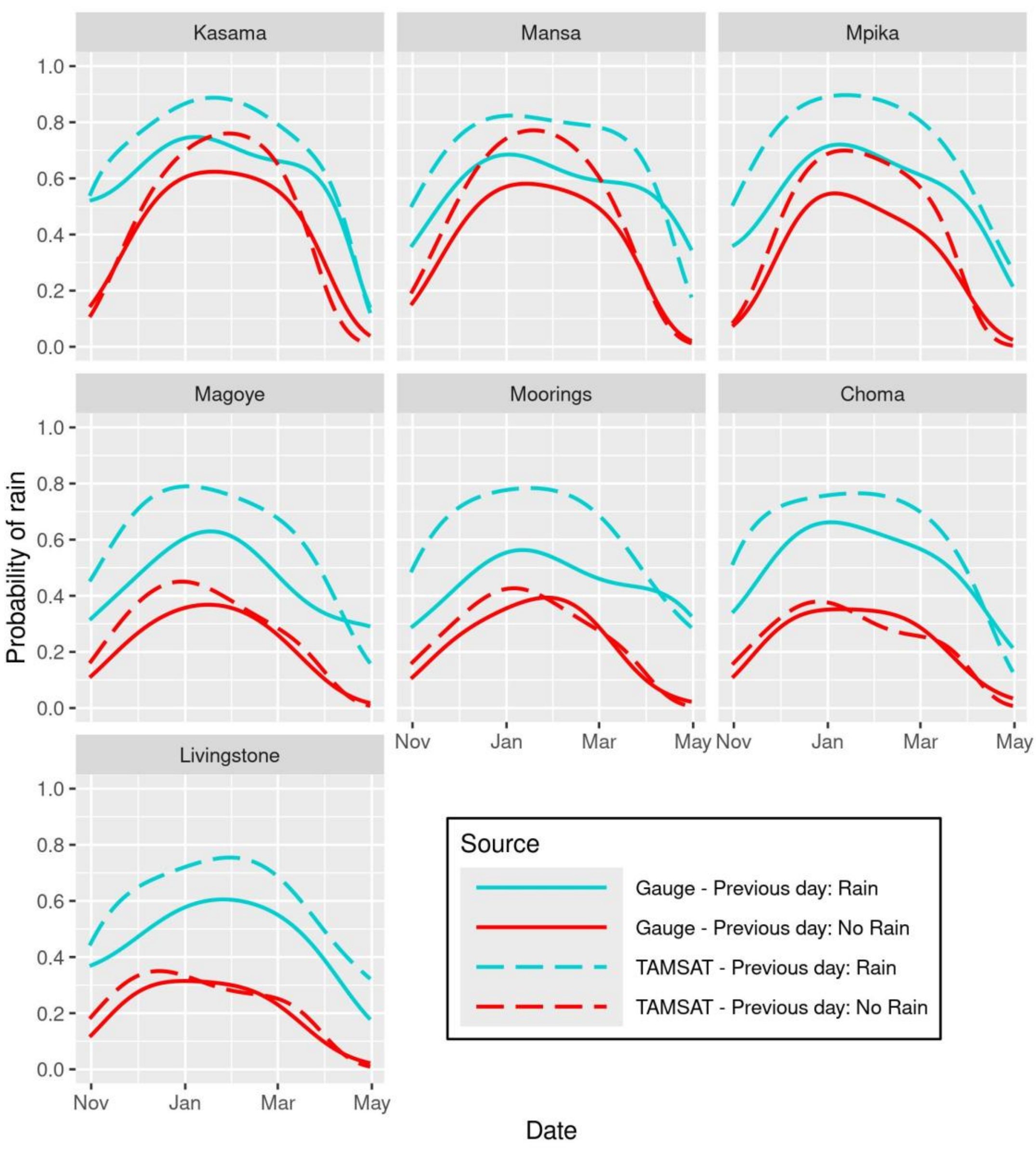

Figure 14: The probability of rain given the previous day being dry or wet for Gauge vs TAMSAT at seven locations in Zambia

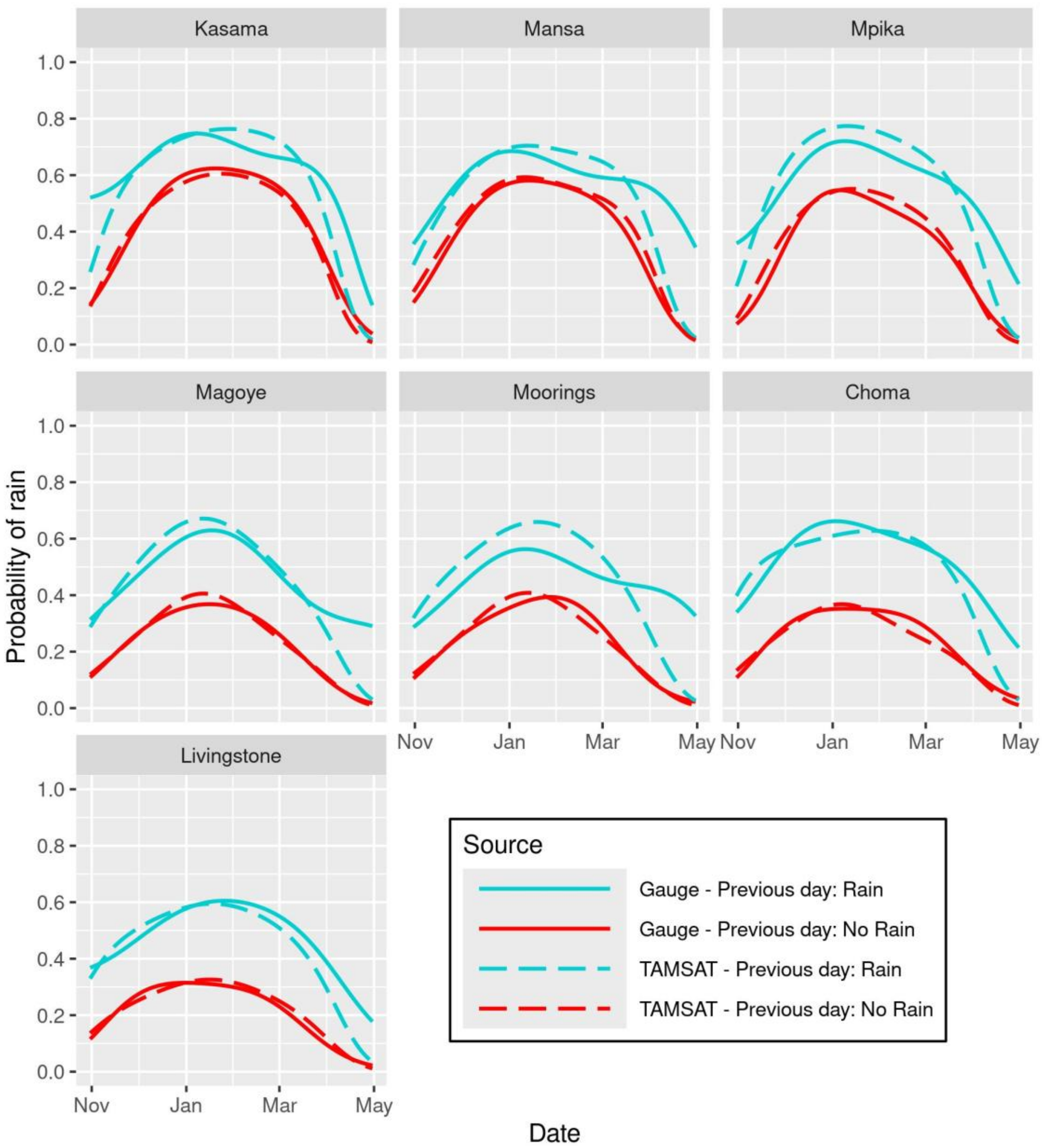

Figure 15: The probability of rain given the previous day being dry or wet for Gauge vs TAMSAT > 4mm threshold at seven locations in Zambia

#### *4.3.3.4 Rainfall amounts and extremes*

The probability of detection for each of the five rainfall amount categories across the seven stations in Zambia is shown in Figure 16 for all SRE. All of the SRE achieve at least approximately 75% probability of detecting  dry days. However the ability to detect other rainfall categories is substantially lower. For example, only TAMSAT and ERA5 achieve a probability of detection >50% for moderate rainfall across all stations. The lowest detection rates for all SRE occur in the violent rain category. Notably, TAMSAT has a probability of detection of 0% and 1% for violent rain. This result indicates a potential limitation of the SRE to detect extreme rainfall events. ENACTS has the most promising overall performance, but it still show relatively low probability of detection of violent rain.

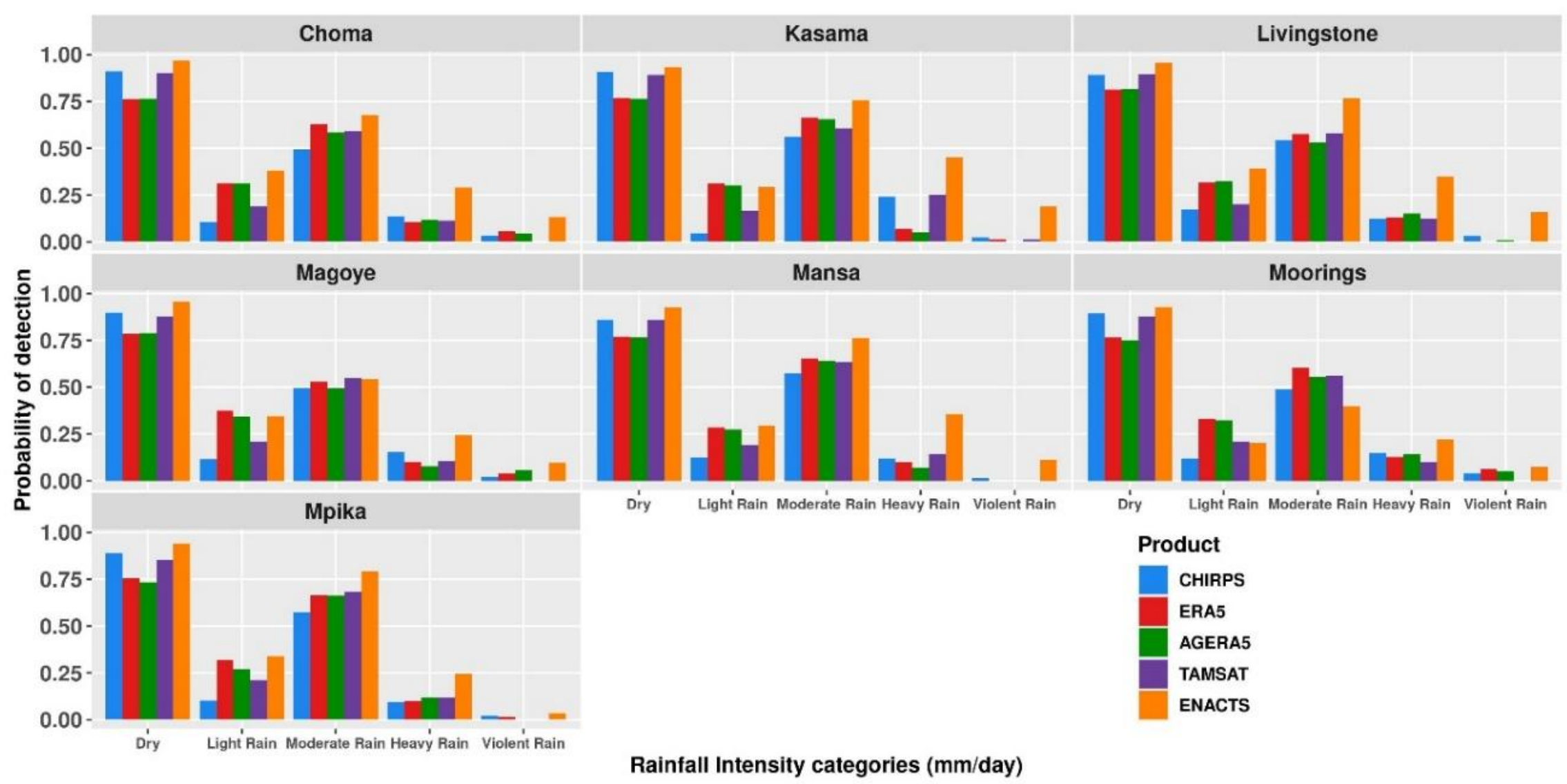

Figure 16: Probability of detection of rainfall intensity categories at seven locations in Zambia

These findings highlight the fact that SRE do not measure the same phenomena as gauge data and may not detect the same type of rainfall on the same day, particularly for high-intensity and extreme rainfall. As a result, SRE may fail to detect the key rainfall events, such as start and end of rains, which are sensitive to small differences in rainfall amounts. These small discrepancies can lead to large differences in event timing, even when total values are close. For example, consider a simple definition of the start of the rains as: "the first date after 1 April when total rainfall over 3 days exceeds 20 mm". If the gauge data satisfies this condition on 15 April, but the corresponding SRE only records 18 mm over the same 3 days, and does not achieve the event until several weeks later, the estimated start of the rains could appear significantly delayed, even though the total rainfall amounts are quite similar.

### *4.3.4 Bias correction of SRE*

The five long term SRE evaluated at the seven stations in Zambia (ERA5, CHIRPS, TAMSAT, ENACTS, and AGERA5) showed relatively small biases in rainfall totals but consistent overestimation in the number of rain days. This suggests that bias correction may be necessary. The frequency of rain day occurrence directly influence key agricultural rainfall summaries used in PICSA, such as the start and end of the rainy season and dry spell lengths, hence bias correction is a justifiable consideration. To address biases in rain day occurrence, calibrated rain day thresholds for each SRE were calculated at each station on a long term monthly basis to match the rain day occurrence of the gauge data. For comparison, thresholds were also derived using data from all locations combined. Due to the low rainfall during the dry season, the months from May to October were treated as a single period. Figure 17 shows the calculated thresholds for each of the periods at each station with a dotted reference line at 0.85mm. The solid black line shows the thresholds over all stations.

All SRE show a clear seasonal effect with a higher threshold in the peak rainy season (December to February). This finding indicates that a season dependent threshold may be required. A moving window approach has also been used by others (Themeßl, et al., 2011), which avoids inhomogeneities at the end of seasonal periods.

ERA5 and AGERA5 consistently require the highest thresholds, indicating a greater tendency to overestimate the number of rain days compared with gauge data. CHIRPS and TAMSAT also show overestimation but to a lesser degree, while ENACTS shows the best agreement with gauge observations, requiring only minor adjustments. Despite some station level variation, the average thresholds across all stations (solid black lines) follow a consistent seasonal pattern, suggesting a location independent threshold may be sufficient. More stations would be needed to confirm this approach.

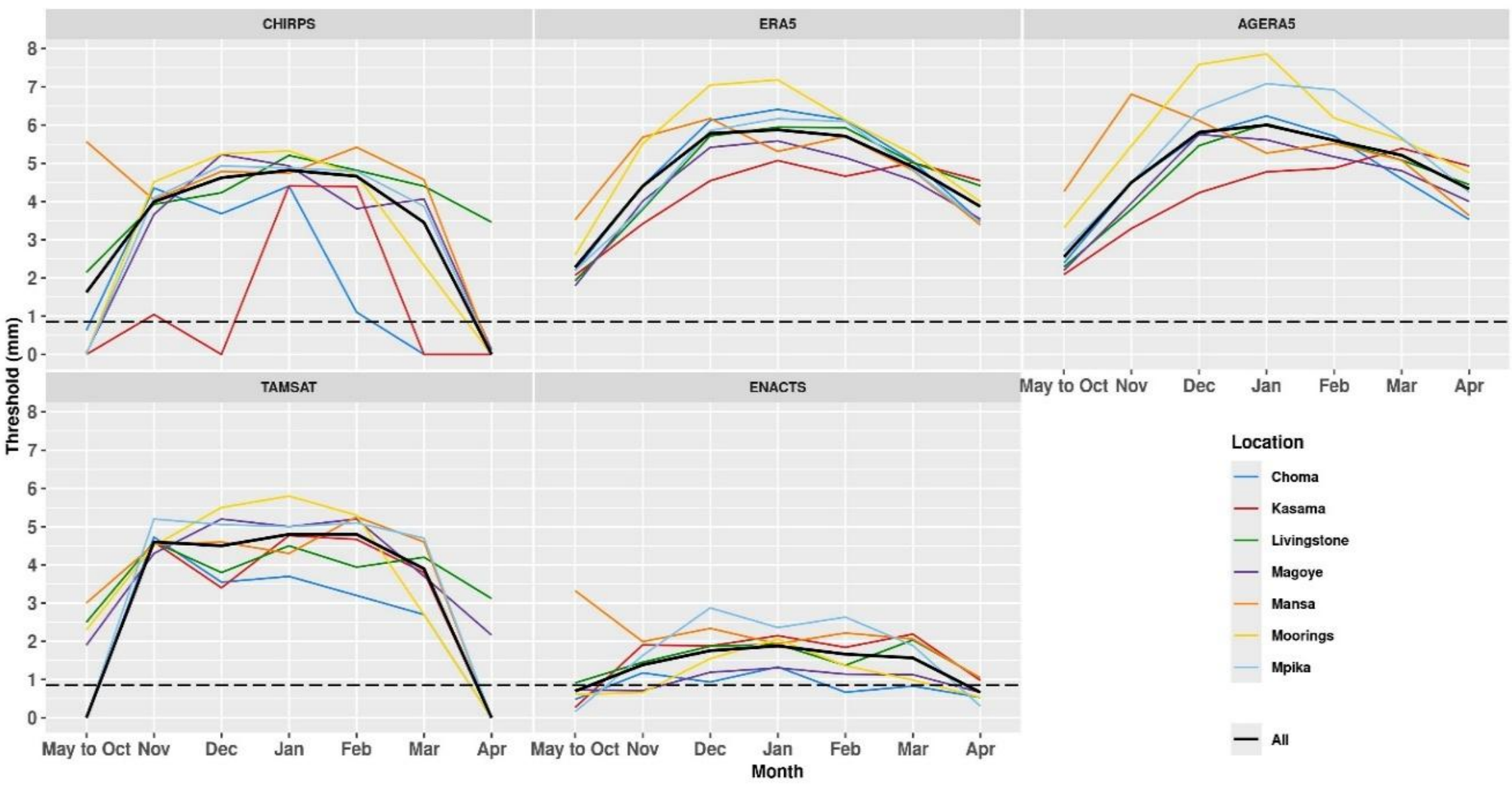

Figure 17: SRE calibrated rain day thresholds for seven locations in Zambia and over all stations together

A detailed evaluation of bias correction methods is not done within this case study and is left for future work. However, the LOCI method was applied to ERA5 data for Zambia, as an illustrative example for use in the next section. Monthly thresholds and scale factors were calculated over all stations combined, with the dry season (May to October) treated as a single period corresponding to the black line in Figure 17. Analysis of the bias corrected values (not shown) by re-running selected comparisons showed that LOCI correction substantially reduced biases in rain day occurrence, along with minor adjustments in rainfall total estimates.

### *4.3.5 Application specific summaries*

A definition for the start of the rainy season in Zambia is the first occasion from 15 November when at least 20mm of rainfall is received over a 3-day period, with no dry spell of 10 days or more in the following 30 days (Stern & Cooper, 2011). Figure 18 presents the start dates for each year at seven stations in Zambia, comparing gauge data with both the uncorrected and LOCI-corrected ERA5 data. Table 10 shows the correlation, mean error and mean absolute error (in days) and the ratio of standard deviations between the gauge data and the two versions of the ERA5 data.

In Choma, Livingstone, Magoye, and Moorings, the LOCI adjusted ERA5 data give a broader range of start dates, and more late onset years, similar to the variability seen in station data, although not always matching the same years. The rSD values indicate that the uncorrected ERA5 values are less variable than those from the station values, while the LOCI correction increases variability, likely due to a reduction in rain day occurrence and therefore a more accurate dry spell distribution. Also, the LOCI correction improves the mean error across the stations, indicating a better long-term average estimate of the start date. However, there is little improvement in the mean absolute error and correlation.

These results suggest that the LOCI-adjusted ERA5 values may be suitable for estimating long term risks of the start date falling above or below certain thresholds (e.g. risk of an early planting date), due to the similar mean and variability observed. However, they may not be reliable for detecting specific historical start of the rain dates in individual years (e.g. identify particularly “good” or “bad” years), due to the low correlation and high mean absolute error. This finding emphasises the importance of evaluating based on the intended application and highlights the need for tailored evaluations focused on practical use cases rather than generic ones.

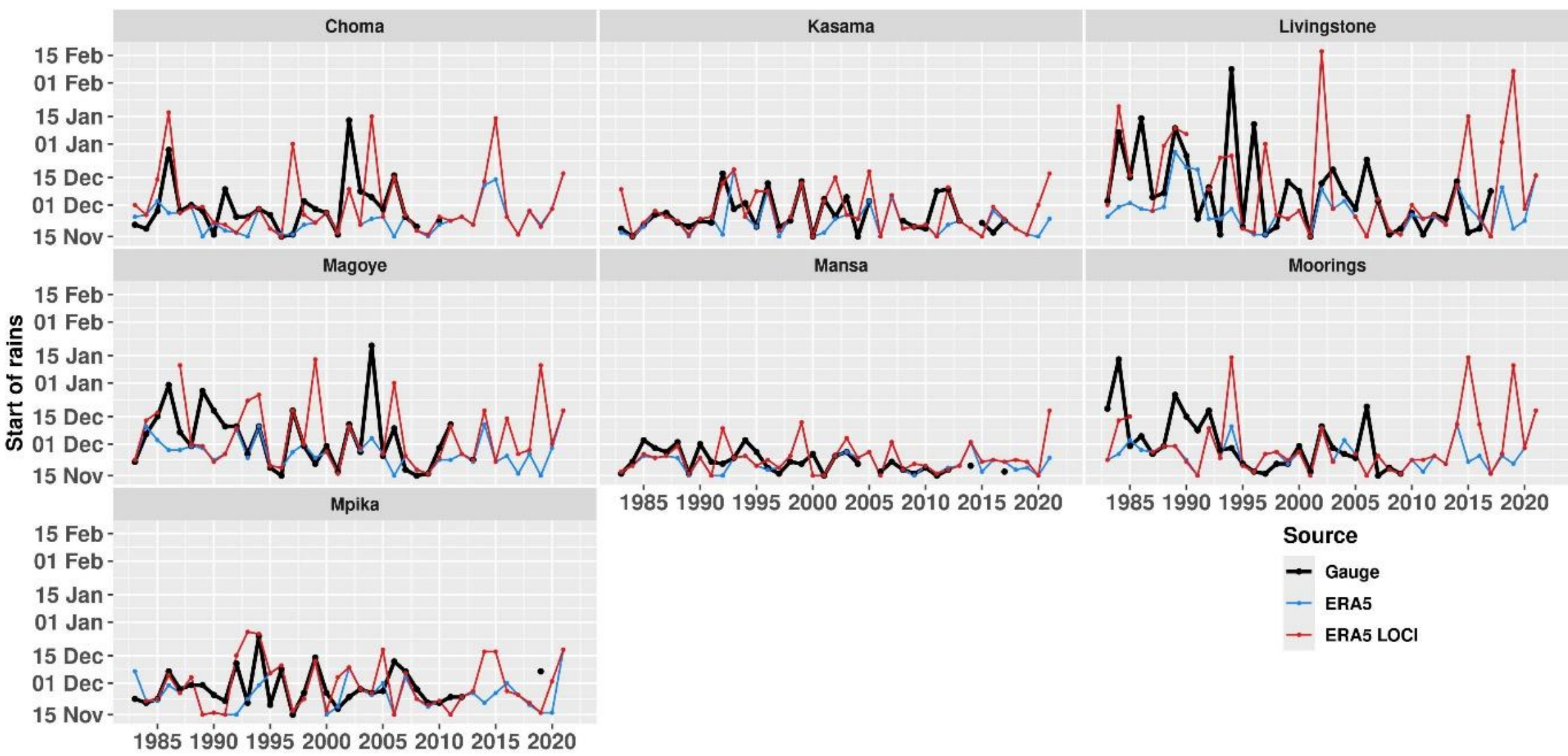

Figure 18: Start of rain dates using a standard definition from Gauge, ERA5 and LOCI-adjusted ERA5 at seven locations in Zambia

Table 10: Comparison metrics for start of rains at seven locations in Zambia. Correlation coefficient (r) and ratio of standard deviations (rSD) are unitless; mean error (ME) and mean absolute error (MAE) are expressed in days.

| Station | Metric | ERA5 | ERA5 LOCI |
|---|---|---|---|
| Choma | r | 0.48 | 0.44 |
| Kasama | r | 0.36 | 0.53 |
| Livingstone | r | 0.33 | 0.31 |
| Magoye | r | 0.51 | 0.39 |
| Mansa | r | 0.39 | 0.37 |
| Moorings | r | 0.24 | 0.27 |
| Mpika | r | 0.24 | 0.39 |
| Choma | ME | -7 | 1.6 |
| Kasama | ME | -2.7 | 1.9 |
| Livingstone | ME | -9.9 | 1.7 |
| Magoye | ME | -7.4 | 3.7 |
| Mansa | ME | -0.5 | 1.1 |
| Moorings | ME | -6.1 | -4 |
| Mpika | ME | -3.7 | 0.2 |
| Choma | MAE | 8.9 | 9.6 |
| Kasama | MAE | 5.9 | 6.2 |
| Livingstone | MAE | 13.7 | 16.2 |
| Magoye | MAE | 9.1 | 9.4 |
| Mansa | MAE | 4.4 | 5 |
| Moorings | MAE | 9.8 | 11.4 |
| Mpika | MAE | 7.7 | 7.7 |
| Choma | rSD | 0.46 | 1.26 |
| Kasama | rSD | 0.92 | 1.19 |
| Livingstone | rSD | 0.5 | 1.16 |
| Magoye | rSD | 0.46 | 1.41 |
| Mansa | rSD | 1.12 | 1.26 |
| Moorings | rSD | 0.5 | 0.87 |
| Mpika | rSD | 0.95 | 1.38 |

# 5 Discussion

## 5.1 Overview

The methodology presented in this paper provides a structured framework for the comprehensive evaluation of SRE for agricultural applications. The case studies demonstrate how this methodology can be used to evaluate aspects of SRE performance that are relevant for agricultural use including strengths, weaknesses, issues, and biases across diverse climates and locations.

The framework begins with an initial data pre-processing step, which involves selecting gauge and SRE data, and temporal and spatial scale choices, based on the application requirements.

The spatial and temporal consistency checks provide an initial screening step, allowing for identification of SRE products with inconsistent spatial or temporal patterns. In the case studies, these checks revealed spatial inconsistencies in some SRE products, likely due to the algorithms used to merge station and satellite data.

The quantitative comparisons enable systematic evaluation of the key components of rainfall data that are known to influence agricultural outcomes. These include annual summary statistics, seasonality (Markov chain models), autocorrelation (spell persistence) and rainfall amount categories and extremes. These analyses expose any inherent biases in the SRE and allowed for investigation of whether these biases vary by season and/or location. In particular, the use of Markov chain models, widely used in rainfall data analysis but not commonly used for SRE comparison, allowed for more detailed investigation of rain day occurrence bias, including its dependence on location, season, and previous day's state (wet or dry). The detection of daily rainfall intensities highlighted a limitation of the SRE data.

These comparisons inform the decision to consider if bias correction is necessary, and comparison of specific summaries can serve as final confirmation of the suitability of SRE. However, biases between gauge and SRE data are to be expected, particularly in rain day occurrence. Several bias correction methods which explicitly address this issue are outlined, and we note that further research is required to understand their suitability for agricultural use. If bias corrections are made, and are shown to improve the general comparisons, then the corrected SRE become candidates for comparison of specific summaries as a final assessment.

## 5.2 Strengths and limitations of the framework

The framework proposed by this paper has several strengths for assessing satellite rainfall estimates for agricultural use. It provides a clear and organised set of steps in a logical sequence, which will help users understand the main aspects of performance that should be examined. This clarity allows for increased transparency in how users may evaluate and select SRE for agricultural applications. Another strength of the framework is that it is adaptable to different applications and can be applied in many different contexts, including with data at different temporal and spatial resolutions making it flexible for a range of applications. It places emphasis on rainfall characteristics that are important for agriculture, including rainfall totals, rainfall occurrence, rainfall event categories and extreme events. The inclusion of an explicit data selection and pre-processing step, and spatial and temporal consistency checks can reveal problems with SRE or gauge data early before more detailed comparisons are carried out, saving time and helping users to be efficient in the evaluation process. While completion of all steps provides the most comprehensive evaluation, when limited data or expertise are available, partial application of the framework can still yield useful insights, allowing users to apply as many steps as are feasible.

There are also limitations. The framework relies on access to gauge data of sufficient quality and duration. In areas with few stations, or where gauge records are inconsistent, the results may be less reliable. Applying the framework also requires technical skills, including statistical programming skills, handling and processing of spatial data files. Obtaining the desired SRE data can also be challenging as access options vary by product. Users without these skills may need additional support, prepared scripts or rely on point-and-click statistical software to use the framework, potentially making it harder to be adapted to a user's specific application. Selecting appropriate bias correction methods also requires expertise and carries risks if not implemented correctly. The temporal checks included in the framework identify clear issues, such as sudden changes, but they do not capture all possible types of temporal problems. Finally, the analyses carried out as part of the methodology may need to be communicated to stakeholder groups and non-technical audiences to justify decisions made about SRE, which is not discussed but could be valuable further work to build into the framework.

## 5.3 Practical considerations to approach under limited resources

Comprehensive implementation of the methodology may not always be feasible in certain operational contexts. In practice, evaluations are often conducted under constraints related to data availability and/or technical capacity. This section outlines how the proposed methodology may be adapted under such constraints and clarifies the implications of these compromises.

Limited gauge data availability can significantly impact the evaluation approach. For applications where spatial evaluation is intended but gauge density is insufficient to generate gridded gauge data, compromises may be needed. There may only be a limited number of locations where gauge data can be transformed to the pixel level, and the spatial interpolation approach may need to be simplified to account for this. In some cases, a spatial evaluation may not be possible and users may need to consider alternative approaches to transform gauge data to a pixel level or accept the mismatch of the spatial representations in the evaluations.

Limited technical expertise or analytical capacity can further constrain how the methodology is applied in practice. In such contexts, it may be appropriate to prioritise SRE products that are straightforward to access and require minimal pre-processing, as data preparation can itself be a substantial barrier. Similarly, if spatial interpolation of the gauge data is required then this may have to rely on simple approaches, such as averaging gauge values within grid cells, rather than more robust but complex methods. Where technical capacity is limited, analyses may need to focus on standard summaries and indices, as calculation of application-specific summaries are often more technically complex to implement. Bias correction, where necessary, may reasonably be restricted to simple approaches, such as mean-based methods. Where bias correction cannot be applied, quantifying and reporting bias in the resulting summaries remains important for appropriate interpretation of SRE performance.

# 6 Conclusion

This methodology complements existing literature on SRE validation and takes a significant step forward in enabling SRE to be evaluated and effectively used to inform agricultural decision making. While general validation studies ask "Does the SRE estimate the true rainfall well?", this methodology further questions "To what extent can an SRE be used for this specific purpose?". The methodology provides a comprehensive framework for evaluating SRE through assessment of core components of rainfall—annual summaries, seasonality, and autocorrelation—while also guiding users in application specific evaluations.

Given the potential consequences of using SRE to support and influence decision making by farmers, there is a need for comprehensive methods to evaluate estimated data for use in agricultural applications. The methodology presented here is intended to support better understanding of SRE evaluation studies and allow a wider community to carry out evaluation studies, supporting greater transparency in the use and understanding of SRE.

This work also contributes to other important applications of SRE. In many countries the network of ground stations includes those with either poor data quality (missing values and suspicious values) or good quality records that are too short for an effective historical analysis. SRE are obvious candidates for infilling missing values, quality controlling values and extending station records, and the methods presented in this paper can contribute to this. For example, the quantitative comparisons can inform which bias correction methods may be suitable for use in infilling missing values.

Finally, although this methodology was developed with agricultural applications in mind, the broader concept of "application specific evaluation" may be equally relevant for other climatic variables and other application domains. The general structure – data preprocessing, spatial and temporal consistency checking, comparisons centred of key rainfall properties relevant to agriculture, consideration and application of bias correction, and lastly the assessment of application-specific summaries – could be considered in a similar way to build on existing general validation studies. This approach could include evaluation of gridded temperature products or SRE for other applications, such as stream flow analysis or producing weather based index insurance, which may have different performance criteria but which could follow this same structure.

# 7 Author Contributions

Danny Parsons: conceptualisation, methodology, analysis, writing – original draft, writing – review and editing.

David Stern: conceptualisation, supervision, writing – review and editing.

Denis Ndanguza: supervision, writing – review and editing.

Mouhamadou Bamba Sylla: supervision, writing – review and editing.

James Musyoka: writing – review and editing.

John Bagiliko: analysis, writing – review and editing.

Graham Clarkson: conceptualisation, writing – review and editing.

Peter Doward: conceptualisation, writing – review and editing.

All authors have read and agreed to the published version of the manuscript.

# 8 Acknowledgments

The authors thank and appreciate the providers of satellite and reanalysis rainfall data, and the Caribbean Institute for Meteorology and Hydrology (CIMH), Ghana Meteorological Agency, Zambia Meteorological Department, and the International Crops Research Institute for the Semi-Arid Tropics (ICRISAT) for providing station rainfall data.

# 9 Funding

This research was funded by a grant from the African Institute for Mathematical Sciences, **www.nexteinstein.org**, with financial support from the Government of Canada, provided through Global Affairs Canada, **www.international.gc.ca**, and the International Development Research Centre, **www.idrc.ca** (accessed on 7 July 2025). This research was also supported by funding from the Oakwell Charitable Trust.

# 10 Conflict of Interest statement

The authors declare no conflicts of interest.

# 11 Data Availability Statement

The satellite and reanalysis data that support the findings of this study are openly available at the Climate Hazards Center in https://data.chc.ucsb.edu/products/CHIRPS-2.0/global_daily/netcdf/p05/ (CHIRPS), Zambia Meteorological Department at http://41.72.104.142/SOURCES/.ZMD/.ENACTS/.rainfall/ (ENACTS), the JASMIN Group in https://gws-access.jasmin.ac.uk/public/tamsat/rfe/data_zipped/v3.1/daily/ (TAMSAT), and the Climate Data Store at https://doi.org/10.24381/cds.adbb2d47 (ERA5) and https://doi.org/10.24381/cds.6c68c9bb (AgERA5). The station data from Barbados, Ghana and Zambia are available from the Caribbean Institute for Meteorology and Hydrology (CIMH), Ghana Meteorological Agency and Zambia Meteorological Department respectively. Restrictions apply to the availability of these data, which were used under license for this study. Data are available from the authors with the permission of the respective organisations. The station data from Niger were provided by the International Crops Research Institute for the Semi-Arid Tropics (ICRISAT) and are available from the corresponding author upon reasonable request.

# 12 Code Availability

Code for data cleaning and analysis associated with this paper is available at https://github.com/dannyparsons/rainfall-comparison-paper (accessed on 7 July 2025).

# 14 List of Figures and Tables